\documentclass[longauth]{tex/src/aa}

\newcommand{\projecttitle}{J-PAS: unprecedented precision in stellar populations of diffuse tidal features}
\newcommand{\projectversion}{0b8f01e}
\newcommand{\projectgitrepo}{https://gitlab.com/sepideh.esk/alba}
\newcommand{\projectcopyrightowner}{Sepideh Eskandarlou <sepideh.eskandarlou@gmail.com>}
\newcommand{\maneagedate}{12 May 2025}
\newcommand{\maneageversion}{a575ef8}

\newcommand{\sedverticalmultip}{19}

\newcommand{\maScaleDec}{43.4945}
\newcommand{\maScaleRAa}{246.735}
\newcommand{\maScaleRAb}{246.741}
\newcommand{\maScaleKpc}{20}

\newcommand{\maCropRAMin}{246.7323976}
\newcommand{\maCropRAMax}{246.7888086}
\newcommand{\maCropDecMin}{43.45556127}
\newcommand{\maCropDecMax}{43.49649686}
\newcommand{\maTickDist}{0.00805871}
\newcommand{\maSBlow}{18}
\newcommand{\maSBhigh}{30}
\newcommand{\maColormap}{gnuastrogray}

\newcommand{\massjpasnum}{6.37}
\newcommand{\massjpasden}{2.68}
\newcommand{\masssdssnum}{6.30}
\newcommand{\masssdssden}{8.98}

\newcommand{\machinearchitecture}{x86\_64}
\newcommand{\machinebyteorder}{Little Endian}

\ifdefined\highlightnew

\else

\fi

\ifdefined\highlightnotes
\newcommand{\tonote}[1]{\textcolor{green!60!black}{[#1]}}
\else
\newcommand{\tonote}[1]{{}}
\fi

\usepackage{graphicx}

\usepackage{fixltx2e}

\usepackage{xcolor}
\color{black} 
\definecolor{DarkBlue}{RGB}{0,0,90}

\usepackage{setspace, caption}
\usepackage[
  colorlinks,
  urlcolor=blue,
  citecolor=blue,
  linkcolor=blue,
  linktocpage]{hyperref}

\hypersetup{
    pdftitle={\projecttitle},
    pdfauthor={\projectcopyrightowner},
    pdfsubject={\projectgitrepo{} (commit \projectversion)},
    pdfkeywords={Reproducible research, Maneage, ADD YOUR OWN}
}

\usepackage{tikz}
\usetikzlibrary{external}
\tikzsetexternalprefix{tex/tikz/}

\usepackage{pgfplots}
\pgfplotsset{compat=newest}
\usepgfplotslibrary{groupplots}
\pgfplotsset{
  axis line style={thick},
  tick style={semithick},
  tick label style = {font=\footnotesize},
  every axis label = {font=\footnotesize},
  legend style = {font=\footnotesize},
  label style = {font=\footnotesize}
  }

\usepackage{txfonts}

\pgfplotsset{
  /pgfplots/colormap={gnuastrogray}{
    rgb255=(0,0,0)
    rgb255=(255,255,255)
  }
}

\title{\projecttitle}
\subtitle{}

\begin{document}

\author{
  Sepideh Eskandarlou\inst{\ref{CEFCA}}
  \and
  Mohammad Akhlaghi\inst{\ref{CEFCA},\ref{CEFCA2}}
  \and
  Francisco Arizo-Borillo\inst{\ref{CEFCA}}
  \and
  Johan H. Knapen\inst{\ref{IAC},\ref{ULL}}
  \and
  Helena Dom\'inguez S\'anchez\inst{\ref{IFCA}}
  \and
  Juan Antonio Fernández-Ontiveros\inst{\ref{CEFCA},\ref{CEFCA2}}
  \and
  Carlos L\'opez-Sanjuan\inst{\ref{CEFCA},\ref{CEFCA2}}
  \and
  Rosa Mar\'ia Gonz\'alez Delgado\inst{\ref{IAA}}
  \and
  Yolanda Jim\'enez Teja\inst{\ref{IAA}}
  \and
  Renato Dupke\inst{\ref{ON}}
  \and
  Yves Revaz\inst{\ref{EPFL}}
  \and
  Pascale Jablonka\inst{\ref{EPFL},\ref{GEPI}}
  \and
  Santi Roca-F\'abrega\inst{\ref{LUND},\ref{MADRID}}
  \and
  Juan Mir\'o-Carretero\inst{\ref{MADRID}, \ref{LEIDEN}}
  \and
  David Mart\'inez-Delgado\inst{\ref{CEFCA},\ref{ARAID}}
  \and
  Alejandro Lumbreras-Calle\inst{\ref{CEFCA}}
  \and
  Antonio Hernán-Caballero\inst{\ref{CEFCA},\ref{CEFCA2}}
  \and
  H\'ector V\'azquez Rami\'o\inst{\ref{CEFCA},\ref{CEFCA2}}
  \and
  Ra\'ul Infante-Sainz\inst{\ref{CEFCA}}
  \and
  Ana L. Chies-Santos\inst{\ref{UFRGS}}
  \and
  Alessandro Ederoclite\inst{\ref{CEFCA},\ref{CEFCA2}}
  \and
  Julio Esteban Rodr\'iguez Mart\'in\inst{\ref{IAA}}
  \and
  Raul Abramo\inst{\ref{USP}}
  \and
  Jailson Alcaniz\inst{\ref{ON}}
  \and
  Narciso Benitez\inst{\ref{IAA}}
  \and
  Silvia Bonoli\inst{\ref{CEFCA},\ref{DICP}}
  \and
  Javier Zaragoza\inst{\ref{CEFCA}}
  \and
  Saulo Carneiro\inst{\ref{ON}}
  \and
  Javier Cenarro\inst{\ref{CEFCA},\ref{CEFCA2}}
  \and
  David Cristóbal-Hornillos\inst{\ref{CEFCA}}
  \and
  Simone Daflon\inst{\ref{ON}}
  \and
  Carlos Hernández–Monteagudo\inst{\ref{IAC},\ref{ULL}}
  \and
  Jifeng Liu\inst{\ref{NAOC}}
  \and
  Antonio Mar\'in Franch\inst{\ref{CEFCA},\ref{CEFCA2}}
  \and
  Claudia Mendes de Oliveira\inst{\ref{USP}}
  \and
  Mariano Moles\inst{\ref{CEFCA}}
  \and
  Fernando Roig\inst{\ref{ON}}
  \and
  Laerte Sodré Jr.\inst{\ref{USP}}
  \and
  Keith Taylor\inst{\ref{Instruments4}}
  \and
  Jesús Varela\inst{\ref{CEFCA}}
  \and
  José Manuel Vilchez\inst{\ref{IAA}}
}

\institute{Centro de Estudios de F\'isica del Cosmos de Arag\'on (CEFCA), Plaza San Juan 1, 44001 Teruel, Spain\label{CEFCA}
  \email{sepideh.eskandarlou@gmail.com}
  \and
  Unidad Asociada CEFCA-IAA, CEFCA, Unidad Asociada al CSIC por el IAA, Plaza San Juan 1, 44001 Teruel, Spain\label{CEFCA2}
  \and
  Instituto de Astrof\'isica de Canarias, Calle V\'ia L\'actea s/n, 38205 La Laguna, Spain\label{IAC}
  \and
  Departamento de Astrof\'isica, Universidad de La Laguna, 38205 La Laguna, Spain\label{ULL}
  \and
  Instituto de Astrof\'isica de Andaluc\'ia (IAA), Consejo Superior de Investigaciones Cient\'ificas (CSIC), Apdo 3004, E-18080, Granada, Spain\label{IAA}
  \and
  Instituto de Física de Cantabria (IFCA), CSIC-Univ. de Cantabria, Avda. los Castros, s/n, E-39005 Santander, Spain\label{IFCA}
  \and
  Observatório Nacional, Ministério da Ciencia, Tecnologia, Inovação e Comunicações, Rua General José Cristino, 77, São Cristóvão, 20921-400, Rio de Janeiro, Brazil\label{ON}
  \and
  Departamento de Astronomia, Instituto de Física, Universidade Federal do Rio Grande do Sul (UFRGS), Av. Bento Gonçalves 9500, Porto Alegre, R.S, Brazil\label{UFRGS}
  \and
  Donostia International Physics Center (DIPC), Manuel Lardizabal Ibilbidea, 4, San Sebastián, Spain\label{DICP}
  \and
  Departamento de Astronomia, Instituto de Astronomia, Geofísica e Ciências Atmosféricas, Universidade de São Paulo, São Paulo, Brazil\label{USP}
  \and
  Instruments4, 4121 Pembury Place, La Canada Flintridge, CA 91011, USA\label{Instruments4}
  \and
  ARAID Foundation, Avda. de Ranillas, 1-D, E-50018 Zaragoza, Spain\label{ARAID}
  \and
  Laboratoire d’Astrophysique, \'Ecole Polytechnique F\'ed\'erale de Lausanne (EPFL), 1290 Sauverny, Switzerland\label{EPFL}
  \and
  GEPI, Observatoire de Paris, CNRS UMR 8111, Université Paris Diderot, 92125 Meudon Cedex, France\label{GEPI}
  \and
  Lund Observatory, Division of Astrophysics, Department of Physics, Lund University, SE-221 00 Lund, Sweden\label{LUND}
  \and
  Departamento de F\'isica de la Tierra y Astrof\'isica, Facultad de Ciencias F\'isicas, Plaza Ciencias, 1, 28040 Madrid, Spain\label{MADRID}
  \and
  Leiden Observatory, Leiden University, J.H. Oort Building, Niels Bohrweg 2, 2333 CA Leiden, Netherlands\label{LEIDEN}
  \and
  National Astronomical Observatory of China,  Chinese Academy of Sciences, Beijing, China\label{NAOC}
}

\date{Received November 30, 2024}

\abstract
    {Galaxies frequently interact with nearby systems, a process that can significantly alter their morphology and star formation activity.
      However, spectroscopic studies of their faint and diffuse remnants require very long exposure times and often exceed the limited field of view of integral field units (IFUs).
      On the other hand, broad-band imaging can have a much wider field of view, but lacks the spectral resolution to identify key spectral features, restricting accurate constraints on stellar population properties.
      With its 54 narrow-band filters in the optical and wide coverage (planned 8000 square degrees), J-PAS fills this gap.
      In this case study, we examine PGC 3087775, a massive galaxy at $z=0.046179$ ($\sim201$ Mpc) in the later stages of a major merger in the J-PAS early data release.
      Photometry was performed on visually-identified polygons (over the tidal features) using Gnuastro and validated with MaNGA IFU data (for the central part).
      Spectral energy distribution (SED) fitting was then carried out with CIGALE to derive stellar population properties using both J-PAS and SDSS photometry to assess the added value with J-PAS.
      SDSS indicates a metal-rich population with an extended star formation history (SFH) and elevated star formation rates.
      J-PAS instead points to a less metal-rich population with moderate extinction and a more rapid SFH, consistent with a quenched stellar population.
      The average $D_n(4000)$ index of the tidal features is 1.24, suggesting that it was a non-dry merger and a fourfold improvement in the precision of stellar mass and $D_n(4000)$ was found with J-PAS.
      We also assessed two heuristic methods for estimating the mass-to-light ratio from SDSS filters and found that they overestimate the stellar mass in this galaxy by 0.5 dex and 0.4 dex relative to SED fitting results from J-PAS and SDSS, respectively.
      Future work will extend this analysis to a larger sample of merging galaxies and evolution of the stellar populations of such structures across the nearby Universe to unprecedented detail.
      This project and its results are fully reproducible, through Maneage (commit \texttt{\projectversion{}}).
      }


\maketitle
\nolinenumbers

\section{Introduction}
\label{sec:introduction}

The prevailing theory of galaxy evolution suggests that galaxies grow through a combination of gas accretion and interactions with other galaxies, including both minor and major mergers \citep{kauffmann93, Cole00, bullock05, stringer07, huillier12}.
Minor mergers induce significant morphological changes in the smaller galaxy; stripping its stars and gas to form tidal structures around the larger galaxy which is much less affected \citep{martinez10, martinezb23}.
In contrast, major mergers typically involve significant perturbations in the interacting galaxies, often accompanied by thick and massive tidal features in comparison to those produced by minor mergers \citep{weilbacher18}.
These tidal features, which appear as low surface brightness (LSB) structures, can serve as tracers of the stellar populations of the progenitor galaxies \citep{halton66, toomre72, Malin80, quinn84, hood18, miro23, miro24a, miro24b}.
A systematic search for tidal stream features, combined with a comprehensive characterization of their stellar populations, including ages, masses, and star formation histories, is crucial for advancing the field, as it sheds light on the hierarchical assembly and evolutionary history of galaxies and provides valuable constraints on the nature of dark matter \citep{fall80, bosch02, agertz11}.

So far, most studies of the photometric properties of tidal remnants have employed broad-band filters \citep{javanmardi16, miro23, laine24, martinez24, miro24a, miro24b} because they offer deep imaging across wide areas.
However, such filters lack the spectral resolution required for precise characterisation of stellar populations.
While spectroscopy can overcome this limitation, observing diffuse tidal structures remains challenging due to the impractically long exposure times needed, even at low resolutions.
Furthermore, the limited fields of view of integral field units (IFUs) restrict the ability to fully map extended features.
For instance, Mapping Nearby Galaxies at APO \citep[MaNGA;][]{bundy15, cherinka19}, which uses the Sloan 2.5m telescope, is only $12-32$ arcsecs in diameter; The PMAS/PPAK at Calar Alto \citep[used in surveys like Calar Alto Legacy Integral Field Area, or CALIFA;][]{sanchez12}, provides an hexagonal shape of $74\times64$ arcsec$^{2}$; and the Multi Unit Spectroscopic Explorer \citep[MUSE;][]{bacon10} at the VLT covers just 59.9$\times60$ arcsec$^{2}$.
These constraints often necessitate multiple pointings for the study of nearby galaxies \citep[for example in PHANGS-MUSE, see][]{phangsmuse} or pre-selection of specific regions within the tidal structures \citep{zhou14, zaragoza18, weilbacher18, zhang20, martinez24}.

The Javalambre Physics of the Accelerating Universe Astro\-physical Survey \citep[J-PAS;][]{benitez14} bridges the gap between broad-band imaging and spectroscopy through its innovative design.
J-PAS employs 54 contiguous narrow-band filters to obtain a low-resolution spectral energy distribution (SED; R = 39–90) for every pixel across its planned 8000 deg$^2$ coverage of the northern sky.
This is achieved without any pre-selection, thereby enabling SED-based selection, or an un-biased SED analysis of extended tidal features for the first time.

\begin{figure}[t]
  \begin{center}
  \ifdefined\makepdf%
    \tikzsetnextfilename{fig-color-img}%
    \input{tex/src/fig-color-img.tex}%
  \else
    \includegraphics[width=\linewidth]{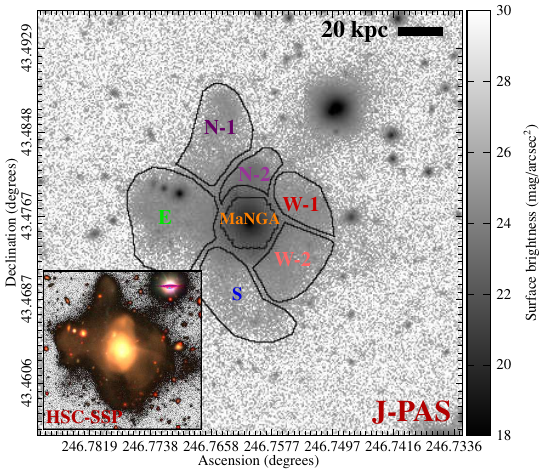}
  \fi

  \end{center}
  \caption{\label{fig:colorimg} Alba galaxy (PGC 3087775) and its major merger remnants.
    The main panel shows surface brightness image of the Alba galaxy that is produced by co-adding all the 57 J-PAS bands (including 54 narrow-band filters, two medium-band filters, and one broad-band filter; however, only the 56 narrow/medium-band filters are used in the analysis presented in this paper).
    The co-add reaches a 3$\sigma$ measured surface brightness limit over 100 arcsec$^2$ of 29 mag/arcsec$^2$.
    The color-bar represents the surface brightness of the co-added image.
    The hexagon over the center of the galaxy delineates the MaNGA footprint used to validate J-PAS SEDs in Fig.~\ref{fig:manga}.
    The six polygons delineated with the thin lines, are named and studied independently in subsequent figures (with same color coding).
    The lower left panel displays a false-color image from the HSC-SSP (Subaru) wide survey, created using the z, r, and g filters as RGB.
    The coordinates of the polygons are available on Zenodo, and links to all corresponding files are provided in Appendix~\ref{appendix:zenodocolorimg}.
  }
\end{figure}

To understand the added advantages of J-PAS for these features, in this work, we use the J-PAS Early Data Release (EDR; V\'azquez Rami\'o et al., in prep.) to investigate the merging galaxy with the brightest tidal features within the J-PAS EDR footprint:  PGC 3087775 (hereafter ``Alba''; see Fig.~\ref{fig:colorimg}).
It is cataloged in the J-PAS EDR as \texttt{9601-18620}, has a spectroscopic redshift of 0.046179 according to NED, based on data from the \citet{desi24}.
Alba is also cataloged as \texttt{PGC 3087775}, \texttt{2MASX J16270254+4328340}, \texttt{SDSS J162702.56+432833.9} and \texttt{MaNGA 1-134964}, among others.
Assuming Planck 2018 cosmology \citep[where $H_0=67.66$ km s$^{-1}$ Mpc$^{-1}$, $\Omega_\Lambda=0.6889$ and $\Omega_m=0.3111$;][]{planckcollab18}, Alba is approximately 201 Mpc away.
Alba is also classified as a Seyfert 2 active galactic nucleus \citep[AGN,][]{toba14} in SIMBAD\footnote{\url{https://simbad.cds.unistra.fr/simbad/sim-id?Ident=PGC+3087775}} \citep{wenger00}.
Alba is characterized by prominent, thick tidal features extending over 56 kilo-parsec (kpc), identifying it as a multi-component system in the advanced stages of a major merger \citep{petersson23}.
This makes Alba an ideal case study for evaluating the consistency of stellar population properties derived from different datasets, specifically J-PAS and SDSS (used in deep imaging surveys like HSC, Legacy Survey or LSST), using various analytical methodologies.

In this paper, we assess the stellar population properties of the tidal features based on their physical criteria, including the spatial configuration and morphology of the tidal structures, as well as the presence of emission or absorption lines within different parts of the merger remnant.
The two primary approaches employed are SED fitting and two heuristic methods for obtaining the stellar mass from SDSS filters: \citet[][hereafter B03]{bell2003} and \citet[][hereafter GB19]{benito19}.
Heuristic methods have been widely adopted in the literature for estimating stellar mass-to-light (M/L) ratios; making them relevant for this work.

This paper is organized as follows.
Sect.~\ref{sec:data} introduces the datasets used and Sect.~\ref{sec:analysis} details the analytical approach and the tools employed for data processing and interpretation.
In Sect.~\ref{sec:results}, we present the results, which are followed by a discussion of their implications in Sect.~\ref{sec:discussion}.
Finally, Sect.~\ref{sec:conclusions} provides a summary of the key findings, highlights the main contributions of the study, and offers concluding remarks.

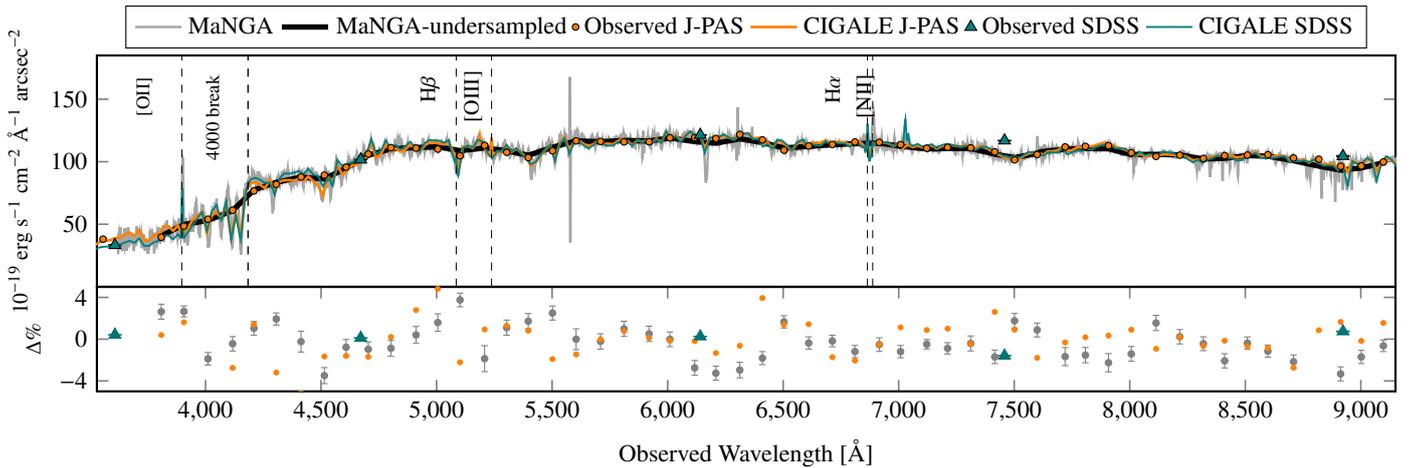
\begin{figure*}[t]
  \begin{center}
  \ifdefined\makepdf%
    \tikzsetnextfilename{fig-manga-jpas}%
\begin{tikzpicture}
  \pgfplotsset{
    every axis legend/.append style={
      at={(0.5,1.03)},
      anchor=south
    },
  }
  \begin{axis}
    [
      xmin=3530,
      xmax=9150,
      ylabel={\tiny 10$^{-\sedverticalmultip}$ erg s$^{-1}$ cm$^{-2}$ {\AA}$^{-1}$ arcsec$^{-2}$},
      width=1\linewidth,
      height=0.25\linewidth,
      enlarge x limits=false,
      enlarge y limits=false,
      scaled ticks=false,
      legend columns=6,
      ytick={50,100,150},
      xtick=\empty,
      tick label style={/pgf/number format/fixed},
    ]

    \addplot [line width=0.9pt, gray!70!white, mark size=2]
    table [x index=0, y index=1] {tex/build/figures/manga-jpas/manga.txt};
    \addlegendentry{MaNGA}

    \addplot [line width=2pt, black!90!black]
    table [x index=0, y index=1] {tex/build/figures/manga-jpas/manga-smooth.txt};
    \addlegendentry{MaNGA-undersampled}

    \addplot [only marks, fill=orange, mark size=1.2, error bars/.cd, y dir=both, y explicit]
    table [x index=0, y index=1, y error index=2] {tex/build/figures/manga-jpas/jpas.txt};
    \addlegendentry{Observed J-PAS}

    \addplot [line width=1pt, orange, mark size=1]
    table [x index=0, y index=1] {tex/build/figures/manga-jpas/cigale-sed-fitting-center-jpas.txt};
    \addlegendentry{CIGALE J-PAS}

    \addplot [only marks, fill=teal, mark=triangle*, mark size=2.5, error bars/.cd, y dir=both, y explicit]
    table [x index=0, y index=1, y error index=2] {tex/build/figures/manga-jpas/sdss.txt};
    \addlegendentry{Observed SDSS}

    \addplot [line width=0.7pt, teal, mark size=2]
    table [x index=0, y index=1] {tex/build/figures/manga-jpas/cigale-sed-fitting-center-sdss.txt};
    \addlegendentry{CIGALE SDSS}

    \addplot[mark=none, black, line width=0.01pt, dashed] coordinates {(3897.42,0) (3897.42,185)};

    \addplot[mark=none, black, line width=0.5pt, dashed] coordinates {(4184,0) (4184,185)};

    \addplot[mark=none, black, line width=0.01pt, dashed] coordinates {(5238.054,0) (5238.054,185)};

    \addplot[mark=none, black, line width=0.01pt, dashed] coordinates {(6864.709,0) (6864.709,185)};

    \addplot[mark=none, black, line width=0.01pt, dashed] coordinates {(6887.48,0) (6887.48,185)};

    \addplot[mark=none, black, line width=0.01pt, dashed] coordinates {(5084.95,0) (5084.95,185)};

  \end{axis}

  \node[anchor=south, rotate=90, scale=0.88] at (0.045\linewidth,0.135\linewidth)
       {\textcolor{black}{\tiny [OII]}};

  \node[anchor=south, rotate=90] at (0.554\linewidth,0.135\linewidth)
       {\textcolor{black}{\tiny [NII]}};

  \node[anchor=south, rotate=90] at (0.53\linewidth,0.14\linewidth)
       {\textcolor{black}{\tiny H$\alpha$}};

  \node[anchor=south, rotate=90] at (0.248\linewidth,0.14\linewidth)
       {\textcolor{black}{\tiny H$\beta$}};

  \node[anchor=south, rotate=90] at (0.278\linewidth,0.135\linewidth)
       {\textcolor{black}{\tiny [OIII]}};

  \node[anchor=south, rotate=90, scale=0.88] at (0.092\linewidth,0.12\linewidth)
       {\textcolor{black}{\tiny 4000 break}};

  \begin{axis}
    [ at={(0\linewidth,-0.0742\linewidth)},
      xmin=3530,
      xmax=9150,
      ymin=-5,
      ymax=5,
     ytick={-4,0,4},
      xlabel={Observed Wavelength [{\AA}]},
     ylabel={\tiny $\Delta\%$},
      width=1\linewidth,
      height=0.16\linewidth,
      enlarge x limits=false,
      enlarge y limits=false,
      scaled ticks=false,
      legend columns=4,
      tick label style={/pgf/number format/fixed},
    ]

    \addplot [only marks, gray, mark size=1.2, error bars/.cd, y dir=both, y explicit]
    table [y expr=\thisrowno{1},
           x expr=\thisrowno{0},
           x index=0, y index=1, y error index=2] {tex/build/figures/manga-jpas/diff-manga-jpas.txt};

    \addplot [only marks, orange, mark size=0.9, error bars/.cd, y dir=both, y explicit]
    table [y expr=\thisrowno{1},
           x expr=\thisrowno{0},
           x index=0, y index=1, y error index=2] {tex/build/figures/manga-jpas/diff-obs-sed-jpas.txt};

    \addplot [only marks, teal!90!black, mark=triangle*, mark size=2.5, error bars/.cd, y dir=both, y explicit]
    table [y expr=\thisrowno{1},
           x expr=\thisrowno{0},
           x index=0, y index=1, y error index=2] {tex/build/figures/manga-jpas/diff-obs-sed-sdss.txt};

  \end{axis}

\end{tikzpicture}%
  \else
    \includegraphics[width=\linewidth]{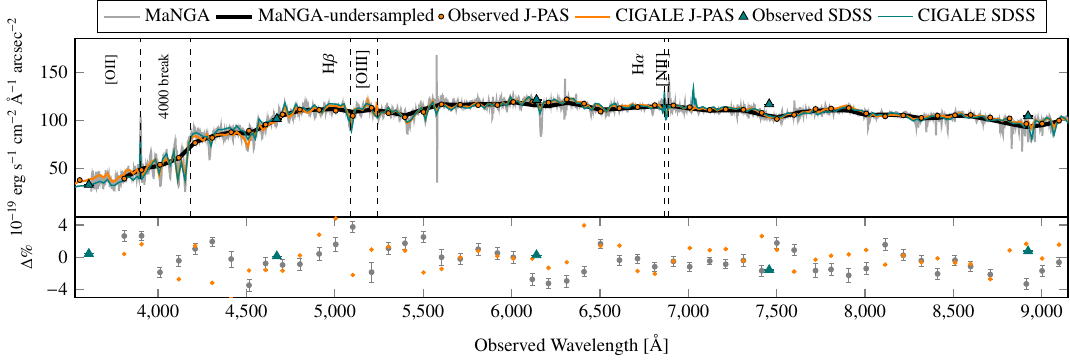}
  \fi

  \end{center}
  \caption{\label{fig:manga}
    Validation of J-PAS SEDs using MaNGA spectroscopy in the central region of the Alba galaxy (shown in Fig.~\ref{fig:colorimg}).
    Top: the grey line indicates the MaNGA spectrum in this area, and the black line shows it after down-sampling with J-PAS filter throughputs.
    The orange points correspond to the measured J-PAS SED, and the orange solid line illustrates the best-fitting SED model derived from J-PAS data in the same region (after applying a minor constant correction factor of $\times1.09$ to all filters).
    Similarly, the teal triangles indicate the measured SDSS SED, and the teal solid line shows the corresponding SDSS SED fit (following the application of a small constant correction factor of $\times0.96$ to all filters).
    Some spectral line positions at the Alba galaxy redshift are also shown.
    Bottom: the percentage difference between the down-sampled MaNGA spectra and the J-PAS photometric data.
    This validates the very good calibration of J-PAS and allows us to study regions beyond the small footprint of MaNGA.
    Furthermore, the orange and teal points and triangles represent the residuals between the SED fits derived from J-PAS and SDSS, respectively, and their corresponding observed fluxes.
    The dataset used to produce this plot can be accessed at Zenodo, see Appendix~\ref{appendix:zenodomanga}.
  }
\end{figure*}

\section{Data}
\label{sec:data}

J-PAS employs an innovative filter system comprising 54 narrow-band filters (14 nm wide), complemented by two medium-band filters (on the extreme blue and red) to encompass the entire optical spectrum from 3216{\AA} to 10839{\AA}, see Fig.~2 of \citet{bonoli21}.
Furthermore, J-PAS also has one broad-band filter (iSDSS) that is used as the detection filter in its main catalog.
The non overlapping J-PAS spectral resolution element (only the narrow-bands) is 10 nm, across the wavelength range from 390 nm to 900 nm.

J-PAS is observed from the Observatorio Astrof\'isico de Javalambre (OAJ) using the Javalambre Survey Telescope (JST250 with a 2.5-meter mirror) that has an impressive effective field of view spanning 7 square degrees (3.5 square degrees exposed).
In this work, we employ images from the J-PAS Early Data Release (EDR), which covers an effective area of $\sim12$ square degrees in all filters.
Details of the dataset like the mean full width at half maximum (FWHM) of the point spread function (PSF), exposure times for each tray and etc, are thoroughly described in  V\'azquez Rami\'o et al. (in prep).

The Alba is detected in the coadded images of all 57 filters.
These coadds combine 482 individual exposures obtained between 29 June 2023 and 6 April 2024.
The observing conditions of the contributing frames cover a broad range, with image quality characterized by a median $\mathrm{FWHM}=0.99''$.
Individual exposure times are 60\,s in all filters, except for the $i{\rm SDSS}$ band where 30\,s exposures are used in order to minimize saturation.
The signal-to-noise ratio (SNR) of the individual detections of the Alba galaxy spans from 21 to 588, with a median value of 185.
As expected, the highest SNR is reached in the $i{\rm SDSS}$ band.
For the narrow-band filters, the SNR follows the combined effect of the instrumental throughput and the spectral energy distribution of the source.

To assess the performance and validate the capabilities of J-PAS, we employ data from the Mapping Nearby Galaxies at APO (MaNGA) survey \citep{bundy15, cherinka19}, which provides spatially resolved spectroscopic data for a large sample of nearby galaxies.
In addition, we utilize imaging data from the Sloan Digital Sky Survey (SDSS) Data Release 9 \citep{blanton17}, which provides observations in five broad-band filters: u, g, r, i, and z.
We have chosen to use SDSS data for several reasons.
Firstly, it offers a depth comparable to that of J-PAS.
Secondly, SDSS (or SDSS-like) filters are used extensively within many other surveys, thereby facilitating comparison and consistency with previous studies.

The Subaru Hyper Suprime-Cam Subaru Strategic Program (HSC-SSP) Data Release 2 \citep{aihara19} are also used as a much deeper dataset to precisely identify and draw the polygons that trace the tidal features and also in Fig.~\ref{fig:colorimg} for a better visual demonstration of Alba.
The 3$\sigma$ surface brightness limit in the HSC-SSP g-band, measured over an area of 100 arcsec$^2$, is 30.31 mag/arcsec$^2$.

\begin{figure*}[t]
  \begin{center}
  \ifdefined\makepdf%
    \tikzsetnextfilename{fig-aperture-photometry-polygon}%
\begin{tikzpicture}

  \pgfplotsset{
    every axis legend/.append style={
      at={(0.5,1.03)},
      anchor=south
    },
  }

  \begin{groupplot}[
      group style={group name=my plots,
                   group size=1 by 2,
                   horizontal sep=0pt,
                   vertical sep=0pt,
                   xticklabels at=edge bottom,
                   yticklabels at=edge left},
      xmin=3500,
      xmax=9580,
      width=1\linewidth,
      height=0.3\linewidth,
      enlarge x limits=false,
      enlarge y limits=false,
      scaled ticks=false,
      legend columns=6,
      tick label style={/pgf/number format/fixed},
      ]

    \nextgroupplot[ylabel={\scriptsize 10$^{-\sedverticalmultip}$ erg s$^{-1}$ cm$^{-2}$ {\AA}$^{-1}$ arcsec$^{-2}$},
                   ymin=0, ymax=10]

    \addplot [line width=1.1pt, violet!80!black, mark size=2, error bars/.cd, y dir=both, y explicit]
    table [x index=0, y index=1, y error index=2] {tex/build/figures/aperture-photometry/flux-density/polygon-1-flux-density.txt};
    \addlegendentry{North-1}

    \addplot [line width=1.1pt, violet!80!white, mark size=2, error bars/.cd, y dir=both, y explicit]
    table [x index=0, y index=1, y error index=2] {tex/build/figures/aperture-photometry/flux-density/polygon-2-flux-density.txt};
    \addlegendentry{North-2}

    \addplot [line width=1.1pt, blue!90!black, mark size=2, error bars/.cd, y dir=both, y explicit]
    table [x index=0, y index=1, y error index=2] {tex/build/figures/aperture-photometry/flux-density/polygon-3-flux-density.txt};
    \addlegendentry{South}

    \addplot [line width=1.1pt, green!90!black, mark size=2, error bars/.cd, y dir=both, y explicit]
    table [x index=0, y index=1, y error index=2] {tex/build/figures/aperture-photometry/flux-density/polygon-4-flux-density.txt};
    \addlegendentry{East}

    \addplot [line width=1.1pt, red!80!black, mark size=2, error bars/.cd, y dir=both, y explicit]
    table [x index=0, y index=1, y error index=2] {tex/build/figures/aperture-photometry/flux-density/polygon-5-flux-density.txt};
    \addlegendentry{West-1}

    \addplot [line width=1.1pt, red!60!white, mark size=2, error bars/.cd, y dir=both, y explicit]
    table [x index=0, y index=1, y error index=2] {tex/build/figures/aperture-photometry/flux-density/polygon-6-flux-density.txt};
    \addlegendentry{West-2}

    \addplot [const plot mark mid, fill=gray, opacity=0.2]
    table [x index=0, y index=2] {tex/build/figures/sb-limit/sb-lim.txt};

    \addplot[mark=none, black, line width=0.5pt, dashed] coordinates {(3813,0) (3813,10)};

    \addplot[mark=none, black, line width=0.5pt, dashed] coordinates {(4184,0) (4184,10)};


    \nextgroupplot[ylabel={\scriptsize ${\rm {mag_{AB}}/arcsec}^{2}$},
                   xlabel={Observed Wavelength [\AA]},
                   ymin=23.05,ymax=27.1,y dir=reverse, legend to name={CommonLegend},legend style={legend columns=4}]

    \addplot [line width=1.1pt, violet!80!black, mark size=2, error bars/.cd, y dir=both, y explicit]
    table [x index=0, y index=1, y error index=2] {tex/build/figures/aperture-photometry/sb/polygon-1-sb.txt};
    \addlegendentry{North-1}

    \addplot [line width=1.1pt, violet!80!white, mark size=2, error bars/.cd, y dir=both, y explicit]
    table [x index=0, y index=1, y error index=2] {tex/build/figures/aperture-photometry/sb/polygon-2-sb.txt};
    \addlegendentry{North-2}

    \addplot [line width=1.1pt, blue!90!black, mark size=2, error bars/.cd, y dir=both, y explicit]
    table [x index=0, y index=1, y error index=2] {tex/build/figures/aperture-photometry/sb/polygon-3-sb.txt};
    \addlegendentry{South}

    \addplot [line width=1.1pt, green!90!black, mark size=2, error bars/.cd, y dir=both, y explicit]
    table [x index=0, y index=1, y error index=2] {tex/build/figures/aperture-photometry/sb/polygon-4-sb.txt};
    \addlegendentry{East}

    \addplot [line width=1.1pt, red!80!black, mark size=2, error bars/.cd, y dir=both, y explicit]
    table [x index=0, y index=1, y error index=2] {tex/build/figures/aperture-photometry/sb/polygon-5-sb.txt};
    \addlegendentry{West-1}

    \addplot [line width=1.1pt, red!60!white, mark size=2, error bars/.cd, y dir=both, y explicit]
    table [x index=0, y index=1, y error index=2] {tex/build/figures/aperture-photometry/sb/polygon-6-sb.txt};
    \addlegendentry{West-2}

    \addplot [const plot mark mid, fill=gray, opacity=0.2]
    table [x index=0, y index=1] {tex/build/figures/sb-limit/sb-lim.txt};

    \addplot[mark=none, black, line width=0.5pt, dashed] coordinates {(3813,23) (3813,27.5)};

    \addplot[mark=none, black, line width=0.5pt, dashed] coordinates {(4184,23) (4184,27.5)};

  \end{groupplot}

  \node[anchor=south, rotate=90] at (0.051\linewidth,-0.043\linewidth)
       {\textcolor{black}{\tiny Balmer break}};

  \node[anchor=south, rotate=90] at (0.107\linewidth,-0.036\linewidth)
       {\textcolor{black}{\tiny 4000 break}};

  \node[anchor=south] at (0.75\linewidth,-0.004\linewidth)
       {\textcolor{gray!70!black}{\footnotesize 3$\sigma$ surface brightness limit over 100 arcsec$^2$}};

  \node[anchor=south] at (0.75\linewidth,-0.22\linewidth)
       {\textcolor{gray!70!black}{\footnotesize 3$\sigma$ surface brightness limit over 100 arcsec$^2$}};

\end{tikzpicture}%
  \else
    \includegraphics[width=\linewidth]{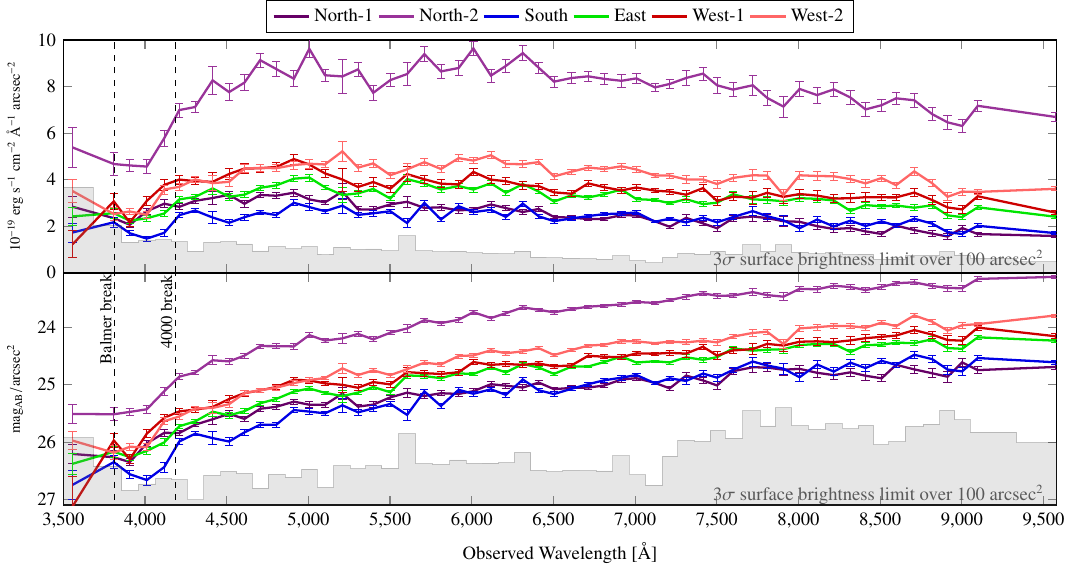}
  \fi

  \end{center}
  \caption{\label{figapen:sedpolygon} The SED (in surface brightness) of different sections of the tidal feature, presented in two units.
    The top panel is in units of wavelength flux density (used in the IFU community) of the six distinct polygons, while the lower panel is in units of AB magnitude (used in the imaging community).
    The colors in this plot correspond to those used in Fig.~\ref{fig:colorimg} to represent different polygons.
    The grey bars at the bottom of each plot show the 3$\sigma$ measured surface brightness limit for each filter over an area of 100 arcsec$^2$.
    The data used to generate this plot is available on Zenodo (see Appendix~\ref{appendix:zenodosedpolygon}).
  }
\end{figure*}
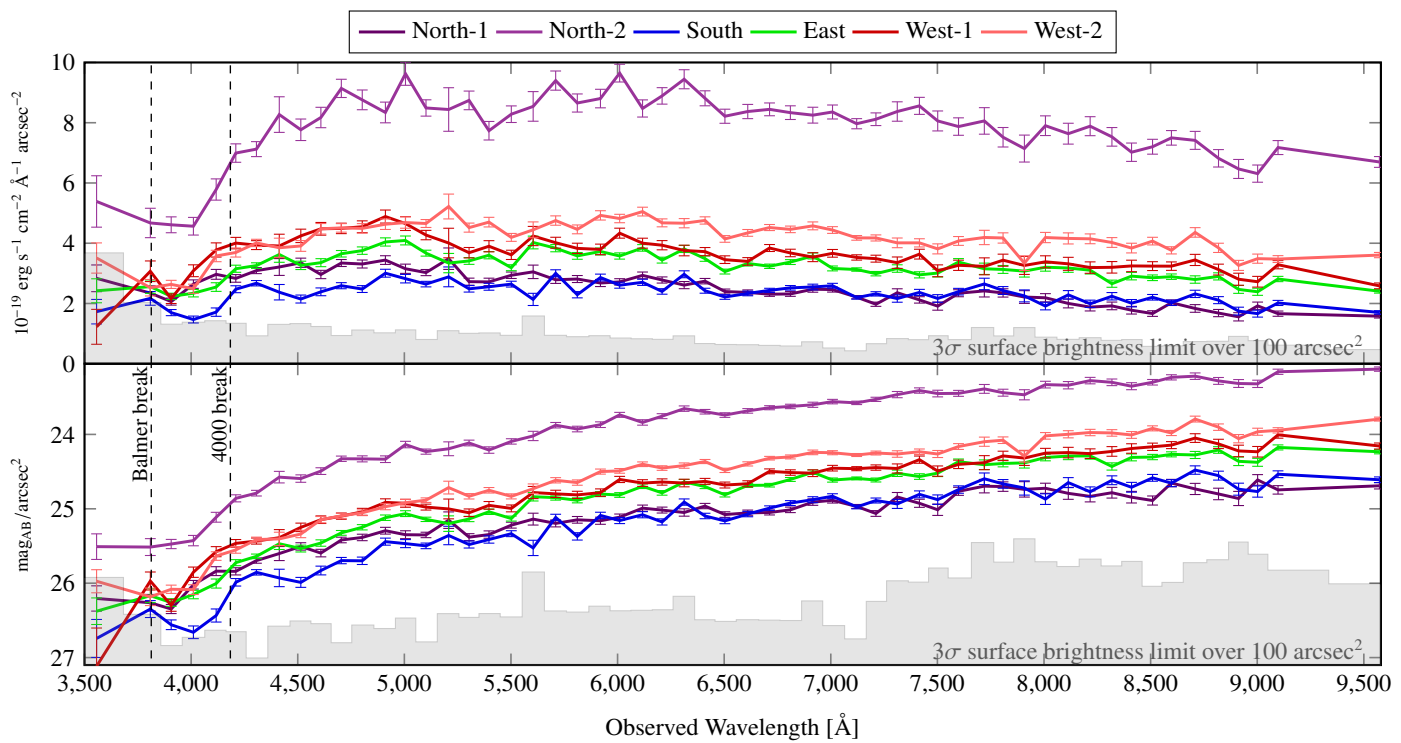

\section{Analysis}
\label{sec:analysis}

\subsection{Photometry and region definitions}
Photometry is conducted using GNU Astronomy Utilities \citep[Gnuastro\footnote{\url{https://www.gnu.org/software/gnuastro}};][]{gnuastrobook}, a comprehensive software suite comprising various programs and library functions designed for the manipulation and analysis of astronomical data.
All programs share a unified command-line interface, enhancing usability for both end users and developers.
Moreover, Gnuastro strictly complies with GNU coding standards, facilitating smooth integration with Unix-based operating systems and existing shell scripts.
A particular Gnuastro program which is important in this analysis is \textsf{MakeCatalog} \citep{akhlaghi19b}, with the executable command \texttt{astmkcatalog}.
This program accepts labeled images to use as apertures of various shapes (non-parametric) and generate a catalog of measurements over each label.

To generate the detection map and effectively separate the signal from noise, we employed \textsf{Noise\-Chisel} \citep{gnuastro} and to mask the foreground and background contaminants, we use the \textsf{Segment} program \citep{akhlaghi19}.
The segmentation map used for masking is then obtained from the co-added image of all 57 J-PAS filters (of which 54 are narrow-band filters, two are medium-band filters, and one is a broad-band filter).
The ``clumps'' of \textsf{Segment} over the polygons in the deeper image (background galaxies or foreground stars) are masked with the same method shown fig.~5 of \citet{martineza23} and we check that no residual flux remained.
The masks were then applied to the photometry in the 56 narrow/medium bands to avoid the contamination of foreground/background sources that are not part of the tidal features.
The PSF of each filter is slightly different, but because we have masked all resolved sources over the streams and their area is much larger than the PSF, the wavelength dependence of the PSF will not affect our analysis \citep[for further details regarding the PSF-subtraction procedure, see][]{Eskandarlou25}.

The differing pixel grids of the different surveys are warped to the pixel grid of J-PAS using the \texttt{-{}-gridfile} option of Gnuastro's \textsf{Warp} program, see Sect.~6.4 of \citet{gnuastrobook}.
In addition, we utilize the \texttt{astscript-color-faint-gray} to generate a color image \citep{infante24}\footnote{Further details on how to display color images of low surface brightness merger remnants are provided in Sect.~5.2.4 ,``Annotations for figure in paper'', of \citet{gnuastrobook}.}.

To identify different regions of the major merging system, we manually draw the named polygons of Fig.~\ref{fig:colorimg} with SAOImage DS9 \citep{ds9}.
They are drawn over the deeper HSC images, based on morphological differences and spatial continuity, in order to effectively separate the various structural components.
Namely, the HSC images are only used to create the labeled images.
The measurements (over those labels) are done on the values of pixels from J-PAS and SDSS using \textsf{MakeCatalog}.
The stellar population properties further described below within Alba’s streams are estimated using this photometry.

The polygons were not generated through an automated image segmentation algorithm in this paper.
This is because any such algorithms will also have systematics that need to be studied in detail, but that is beyond the scope of this work (on the stellar population properties of a new filter-set rather than on the methodological optimization of stream extraction algorithms).
A comparison between visually defined regions and those identified using automated segmentation algorithms constitutes an important avenue for future work on a larger sample of galaxies.

\subsection{J-PAS comparison with MaNGA}

To validate J-PAS calibration and our photometry, Fig.~\ref{fig:manga} compares the J-PAS photometry with MaNGA spectroscopy over the MaNGA footprint shown in Fig.~\ref{fig:colorimg}.
To further assess the reliability of J-PAS measurements, we down-sampled the MaNGA spectrum using the J-PAS filter set.
In the literature, this down-sampling is sometimes called ``convolution'', because applying the filter transmission curve to a spectra has some similarities to a convolution kernel.
But this term can lead to confusion because convolution does not alter the sampling; it only changes the resolution (the number of elements remains the same).
Furthermore, in convolution the kernel moves, but it is fixed in this usage.

The MaNGA spectrum in Fig.~\ref{fig:manga} lacks the typical AGN emission lines, such as [O III]; instead, only [O II] and [N II] are detected, leading to its classification as a low-ionization nuclear emission-line region \citep[LINER;][]{alban23} galaxy.
No significant differences are observed between J-PAS and the down sampled MaNGA data in Fig.~\ref{fig:manga}.
Furthermore, Fig.~\ref{fig:manga} presents a comparison between the SED fitting results derived from J-PAS and those obtained from SDSS; this comparison is discussed in greater detail in Section.~\ref{subsec:sedfitting}.

\begin{figure*}[t]
  \begin{center}
  \ifdefined\makepdf%
    \tikzsetnextfilename{fig-cgal}%
    \input{tex/src/fig-cgal.tex}%
  \else
    \includegraphics[width=\linewidth]{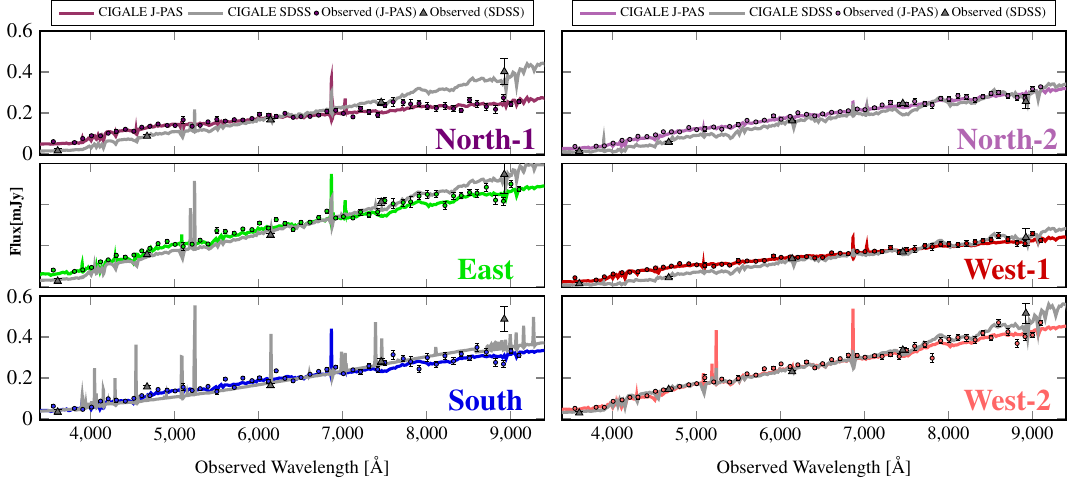}
  \fi

  \end{center}
  \caption{\label{fig:fitcigale}
    SED fitting on Alba's tidal features.
    The measured SEDs and fitted spectra of the left and right panels show the inputs and outputs of CIGALE for both J-PAS and SDSS.
    They correspond to the named polygons of Fig.~\ref{fig:colorimg} with same color.
    Circles show fluxes from 56 narrow/medium-band J-PAS filters, while triangles represent SDSS’s 5 broad filters.
    Fitted CIGALE spectra are overlaid using gray (SDSS) and colored (J-PAS) lines.
    The residuals between the SED fitting results of both surveys and their respective observed fluxes are presented in Appendix.~\ref{appendix:residual} (Fig.~\ref{fig:figresidual}).
    The dataset underlying this plot is publicly accessible through Zenodo (see Appendix~\ref{appendix:zenodofitcigal}).
    }
\end{figure*}

The minor differences between the down-sampled MaNGA and J-PAS occur on absorption lines.
The MaNGA observations (April 2015) and J-PAS data (July 2023) are separated by a substantial temporal gap.
But the differences cannot be attributed to temporal variability of AGNs.
This is because variability is a behaviour characteristic of Seyfert 1 nuclei due to changes in the broad-line region, not Seyfert 2 (the case for Alba).
Furthermore, the absorption in H$\beta$ shows the strongest difference between MaNGA and J-PAS.
But H$\beta$ originates from the underlying stellar population rather than the AGN, and is not expected to have variability on this time scale.
The difference on absorption lines is therefore most plausibly explained by measurement uncertainties rather than intrinsic variability.
A global scaling factor of 1.09 was applied to the J-PAS fluxes, while a factor of 0.96 was used for the SDSS fluxes in order to place both datasets on the same absolute scale as the MaNGA measurements.
  Such re-scaling is a common procedure when combining photometric and spectroscopic observations, as systematic differences frequently arise from variations in aperture size, calibration methods, and the full width at half maximum (FWHM) between the respective instruments and observational techniques \citep[e.g.,][]{Logro19}.
  The most significant outcome of this procedure is the strong agreement in the resulting spectral energy distribution (SED) shape.
  Specifically, the comparison indicates that the SED derived from J-PAS and SDSS photometry is consistent with the spectroscopic measurements at approximately the 4\% level, demonstrating a high degree of reliability in the cross-calibration between the datasets.

Having validated the J-PAS photometry on the overlapping region with MaNGA, we now extend our analysis to the outer tidal features, where no IFU spectroscopy is available.
Fig.~\ref{figapen:sedpolygon} shows the J-PAS photometry (in units of surface brightness) of different polygons over the tidal features from Fig.~\ref{fig:colorimg}, in both wavelength flux density (top) and AB magnitudes (bottom).
Recall that the AB-magnitude system is based on the frequency flux density.
It is important to show both units, as the two communities (IFU and wide-field broad-band imaging) use different conventions and the conversion is not linear.
From the bluer to the redder filters, the slope of the wavelength flux density increases, whereas the slope of the AB magnitude decreases.
This occurs because of the inverse relationship between wavelength and frequency in the two units.
Compared to Fig.~\ref{fig:manga}, the S/N in the center of Alba is 20 times higher than in the tidal streams, hence the error bars are more visible here.

Fig.~\ref{figapen:sedpolygon} also shows the measured surface brightness limit \citep[mSBL, defined in Sect.~7.4.5.1 of][]{gnuastrobook}, represented by grey bars.
The mSBL of each filter is shown in the two units for a general understanding of J-PAS capabilities regarding the low surface brightness regime, not in particular for Alba.
For example, in the two bluest filters, the measured flux of some of Alba's tidal features become fainter than the mSBL.
But this is not a cause for concern because the polygon areas are considerably larger than the 100 arcsec$^2$ used to estimate the mSBL, with the smallest area being at least three times greater.
Furthermore, the mSBL is measured on the noise ($\sim\mkern-8.45mu\sqrt{A}$; where $A$ is the area in number of pixels), while the fluxes are measured on signal ($\sim\mkern-5mu{A}$).
Ultimately, later processing (SED fitting) accounts for the different error bars of each filter, so the larger error bars of the two bluest filters will not have a significant impact on the conclusions of this paper.
In contrast, the mSBL in the red filters (redder than 7200\AA) is brighter due to the presence of telluric emission lines from the atmosphere, which introduce increased Poisson noise.

Additionally, in Fig.~\ref{figapen:sedpolygon}, the North-2 polygon appears significantly elevated relative to the other tidal streams.
This is consistent with the color image in Fig.~\ref{fig:colorimg}, where North-2 is visually brighter than the other regions and has a smaller area, confirming its higher surface brightness.

\subsection{SED fitting}
\label{subsec:sedfitting}
We employ the Code Investigating GALaxy Emission \citep[CIGALE, version 2022.0;][]{boquien19} to perform SED fitting in each of the polygon regions defined in Fig.~\ref{fig:colorimg}.
CIGALE is grounded in the principles of energy balance and utilizes a Bayesian-like statistical framework.
CIGALE is specifically designed to investigate galaxy evolution by comparing modeled SEDs with observed data across a broad wavelength range, from X-rays and far-ultraviolet to far-infrared and radio frequencies.
An important feature of CIGALE for us is its ability to simultaneously model both stellar continuum emission and ionized gas (emission lines).
Although Alba’s tidal features do not show strong emission lines, this capability will be crucial for future studies of merging galaxies in J-PAS where the outskirts may contain significant emission from stripped gas, circumgalactic inflows, or nuclear outflows.
J-PAS is well suited to detect such features in the local Universe.

The input to CIGALE for each polygon consists of flux measurements and their associated uncertainties, derived from the 56 medium/narrow J-PAS bands (the broad-band filter is excluded in the SED fitting) and SDSS (5 broad-band filters), represented as colored circles and gray triangles, respectively, in Fig.~\ref{fig:fitcigale}.
The resulting CIGALE-fitted spectra are illustrated as solid lines in Fig.~\ref{fig:fitcigale}.
The spectra and circles show the J-PAS-derived results and follow the same color coding of the polygons in Fig.~\ref{fig:colorimg}.
The gray spectra and triangles are derived from SDSS.
Additionally, Fig.~\ref{fig:manga} presents both the flux measurements and the corresponding SED-fitting outputs from CIGALE: the J-PAS results are indicated by orange dots and a solid orange line, whereas the SDSS results are shown using teal triangles and a solid teal line.

The stellar population parameters derived by CIGALE are provided in two forms: ``best-fit'' estimates and ``Bayesian'' values, the latter is accompanied by corresponding uncertainties.
In our analysis, we employed the ``Bayesian'' values.
For our SED fitting, we adopt libraries of single stellar populations (SSPs) from the BC03 module \citep{bruzual03} and used the \texttt{sfhdelayed} module.
This module defines an analytic star formation history (SFH) based on a delayed model that may include an optional exponential burst (which could have been induced by the merger); assuming a \citet{chabrier03} initial mass function (IMF).
In addition, we also ran CIGALE with the \citet{salpeter55} IMF for later comparison with heuristic M/L methods (discussed in Sec.\ref{sec:selfconsistency} and Appendix~\ref{appendix:salpeter}) and with a disabled burst and nebular emission for assessing their impact on the results below (see Appendix~\ref{appendix:nostarburst}).

To set the CIGALE parameters (see Appendix~\ref{appendix:cigaleconfigure}), we used an iterative approach.
At first, we used broad \texttt{age\_main} bins and then refined them according to the first results.
For instance, the early runs showed that the age-main value was between 4000 and 5500 Myr.
After that, we narrowed the bins around this range and gradually increased the time resolution until the error bars became stable and the \texttt{age\_main} value remained unchanged.
This systematic adjustment of the input parameters ensured consistency between the Bayesian and the best-fit results.

\begin{table*}[t]
  \caption{CIGALE best-fit parameters. The data used to generate this table are available on Zenodo (see Appendix~\ref{appendix:zenodotabels}).}
\label{tbl:cigalout}
\small
\centering
\resizebox{2.1\columnwidth}{!}{
\begin{tabular}{lccccccc}
\hline\hline
Properties & North-1 & North-2 & South & East & West-1 & West-2 & Center \\
\hline\hline
$J: D_n{4000}$    &    1.08±0.05    &    1.35±0.06    &    1.28±0.05    &    1.28±0.05    &    1.20±0.07    &    1.30±0.06    &    1.46±0.06 \\
$S: D_n{4000}$    &    1.35±0.21    &    1.47±0.27    &    1.08±0.12    &    1.27±0.18    &    1.49±0.26    &    1.26±0.16    &    1.43±0.19 \\ \hline
$J: log (M/M_{\odot})$    &    9.06±0.01    &    9.28±0.01    &    9.27±0.01    &    9.44±0.02    &    9.06±0.01    &    9.43±0.02    &    10.76±0.31 \\
$S: log (M/M_{\odot})$    &    9.53±0.04    &    9.49±0.03    &    9.36±0.04    &    9.63±0.06    &    9.42±0.03    &    9.53±0.04    &    10.83±0.56 \\ \hline
$J: Age_{Burst} (Gyr)$    &    0.34±0.20    &    0.28±0.20    &    0.19±0.18    &    0.16±0.16    &    0.41±0.17    &    0.20±0.19    &    0.40±0.17 \\
$S: Age_{Burst} (Gyr)$    &    0.30±0.21    &    0.34±0.20    &    0.15±0.20    &    0.30±0.21    &    0.32±0.21    &    0.31±0.21    &    0.40±0.17 \\ \hline
$J: Age_{Main}(Gyr)$    &    4.66±0.69    &    4.65±0.58    &    4.75±0.52    &    4.74±0.53    &    4.56±0.95    &    4.74±0.54    &    4.54±0.74 \\
$S: Age_{Main}(Gyr)$    &    4.69±0.62    &    4.71±0.58    &    4.52±0.91    &    4.64±0.69    &    4.72±0.57    &    4.61±0.76    &    4.62±0.77 \\ \hline
$J: Z (\times 10^{-2})$    &    0.28±0.68    &    0.77±0.69    &    0.86±0.80    &    0.89±0.77    &    1.05±1.09    &    1.10±1.07    &    1.17±0.96 \\
$S: Z (\times 10^{-2})$    &    1.66±1.79    &    1.76±1.90    &    1.35±1.77    &    1.23±1.58    &    2.13±1.95    &    1.26±1.58    &    1.59±1.53 \\ \hline
$J: F_{Burst} (\times 10^{-1})$    &    0.53±0.84    &    0.07±0.18    &    0.06±0.10    &    0.07±0.12    &    1.79±0.60    &    0.07±0.15    &    0.08±0.21 \\
$S: F_{Burst} (\times 10^{-1})$    &    0.27±0.62    &    0.10±0.33    &    0.52±0.83    &    0.38±0.73    &    0.10±0.32    &    0.49±0.81    &    0.46±0.80 \\ \hline
$J: E(B-V)_{s} (mag)$    &    0.03±0.06    &    0.05±0.04    &    0.04±0.04    &    0.03±0.04    &    0.16±0.09    &    0.07±0.05    &    0.10±0.04 \\
$S: E(B-V)_{s} (mag)$    &    0.24±0.11    &    0.22±0.12    &    0.23±0.12    &    0.22±0.11    &    0.23±0.11    &    0.20±0.11    &    0.18±0.12 \\ \hline
$J: SFR (M_{\odot}yr^{-1})$    &    0.13±0.28    &    0.07±0.18    &    0.26±0.40    &    0.28±0.51    &    0.14±0.35    &    0.42±0.64    &    1.35±3.46 \\
$S: SFR (M_{\odot}yr^{-1})$    &    1.09±1.94    &    0.35±0.99    &    9.14±12.31    &    1.92±3.69    &    0.44±1.07    &    1.49±2.86    &    5.82±15.69 \\ \hline
$J: \tau_{Burst}$    &    28.28±34.67    &    24.32±31.65    &    25.18±31.31    &    17.53±24.63    &    24.93±31.35    &    24.57±30.91    &    29.55±34.78 \\
$S: \tau_{Burst}$    &    32.87±35.82    &    31.31±35.16    &    37.78±37.07    &    33.68±36.20    &    31.77±35.42    &    33.27±36.15    &    30.54±34.65 \\ \hline
$J: \tau_{Main}$    &    311.44±325.04    &    253.37±233.14    &    341.85±331.15    &    301.33±306.00    &    291.85±307.10    &    372.36±343.03    &    246.57±218.75 \\
$S: \tau_{Main}$    &    379.47±364.15    &    251.97±269.96    &    497.00±406.20    &    421.54±382.35    &    275.64±293.67    &    438.51±387.68    &    276.07±285.40 \\ \hline

\end{tabular}}
\end{table*}

We explore all the remaining available CIGALE configuration options and carefully adjust\footnote{The CIGALE input parameters and output properties are publicly available on SoftwareHeritage as \href{https://archive.softwareheritage.org/swh:1:dir:7c203a565ec6ef40d8419dbdf70e8639b78f4b99;origin=https://gitlab.com/Sepideh.Esk/alba;visit=swh:1:snp:adc584ff3540ea3151de6ebc8462805e8818b1f4;anchor=swh:1:rev:f9d0f9cb25c80415107c962cc53d369a650e9b82}{\texttt{swh:1:dir:7c203a56\-5ec6ef40\-d8419dbdf\-70e8639b7\-8f4b99}} and \href{https://doi.org/10.5281/zenodo.17674805}{zenodo.17674805}, respectively.} the input parameters in incremental steps to achieve the best fit.
The grid parameters that were modified in this analysis are listed in Appendix.~\ref{appendix:cigaleconfigure}.
The output properties reported in Table~\ref{tbl:cigalout} correspond to the following CIGALE output columns: $E(B-V)_{s}$ (\texttt{bayes\-.atten\-uation\-.E-BVs}; $E(B-V)_{s}$, the color excess of the stellar light for both the young and old population), $D_n(4000)$ (\texttt{bayes\-.param\-.D-4000}; D4000 break using the \citep{balogh99} definition), Age Burst (\texttt{bayes\-.sfh\-.age\--burst}; age of the late burst), Age Main (\texttt{bayes\-.sfh\-.age\--main}; age of the main stellar population in the galaxy), Fraction Burst (\texttt{bayes\-.sfh\-.f\--burst}; mass fraction of the late burst population), Z (\texttt{bayes\-.stellar\-.metallicity}; metallicity), SFR (\texttt{bayes\-.sfh\-.sfr}; instantaneous SFR), $M_\star$ (\texttt{bayes\-.stellar\-.m\--star}; total stellar mass), $\tau_{Burst}$ (\texttt{bayes\-.sfh.\-tau\--burst}; duration (rectangle) or e-folding time of all short events), and $\tau_{Main}$ (\texttt{bayes\-.sfh.\-tau\--main}; e-folding time of the main stellar population model).
A comprehensive description of these output parameters and their physical interpretation is provided by \citep{boquien19}.

To assess the reliability of the CIGALE errors, we introduce Gaussian noise with a standard deviation of $1\sigma$ (where $\sigma$ represents the error in measured flux from \textsf{MakeCatalog}).
This process is repeated 100 times for each polygon, implemented via the \texttt{mknoise-sigma} operator of Gnuastro's \textsf{Arithmetic} program.
Since we study seven different polygons (including the central region) the procedure resulted in a large table containing 700 rows of simulated error values.
We then run CIGALE for these 700 random flux errors using both J-PAS and SDSS data.
The results suggest that the error bars provided by CIGALE are accurate.
For instance, in the case of the East polygon, the stellar mass derived from the 700 random flux variations was 9.44±0.02 for J-PAS and 9.62±0.05 for SDSS.
These results are consistent with the error bars directly provided by CIGALE (see Table~\ref{tbl:cigalout}), which report 9.44±0.02 for J-PAS and $9.63\pm0.06$ for SDSS.

For some of the merger remnants, the SDSS fluxes and fitted SEDs in the blue filters are lower than those obtained from J-PAS, whereas in the redder filters they can be higher.
This discrepancy could arise from photometric reduction issues affecting specific SDSS filters, although the possibility of an underlying systematic effect cannot be entirely excluded.
Importantly, this behaviour is not observed in all polygons.
Given the substantially larger number of filters employed by J-PAS, the impact of potential photometric systematics in any single filter is significantly less than SDSS.
Consequently, systematic effects in individual filters are less likely to introduce pronounced biases in the reconstructed SED compared to analyses based solely on SDSS data.

A noticeable difference between the SED fittings derived from J-PAS and SDSS fluxes is that the SDSS fitting for the East and South regions display numerous emission lines, that are not visible in J-PAS fitting.
Given that the SDSS SED fitting reveals such features, while Alba does not demonstrate such strong emission or absorption lines in J-PAS photometry in the merger remnants (even in the central regions of the galaxies), the presence of so many emission lines in the SDSS data highlights the need for further investigation.
Therefore, we reran CIGALE with the starburst and nebular emission components disabled (see Appendix~\ref{appendix:nostarburst}).
We found that, after removing these components, the CIGALE output properties studied (for the stellar population, not gas or emission lines) remained largely unaffected.
This outcome is most likely attributable to the stellar populations in the tidal features and the lower spectral resolution of SDSS.
As the extra emission lines in the two polygons do affect the central argument of this paper (primarily concerned with stellar populations, not gas emission), all subsequent CIGALE analyses were conducted with the starburst and nebular emission components enabled.

\section{Results}
\label{sec:results}

A unique feature of J-PAS is its ability to identify galaxies and their tidal structures based on spectral characteristics and stellar population properties without the need for pre-selection, operating in a fully blind manner.
This paper serves as a proof-of-concept case study that will later be expanded to encompass a larger sample of galaxies.
In the following subsections, particular emphasis is placed on the stellar populations, comparing the results obtained from J-PAS with those derived from SDSS, followed by a study of the self-consistency of the SED fitting results between J-PAS and SDSS.
Finally, we focus on stellar masses with a comparison of the results from SED fitting with two heuristic methods that are commonly used in the literature (with broad-band photometry).

\begin{figure*}[t]
  \begin{center}
  \ifdefined\makepdf%
    \tikzsetnextfilename{fig-corner-plot}%
    \input{tex/src/fig-corner-plot.tex}%
  \else
    \includegraphics[width=\linewidth]{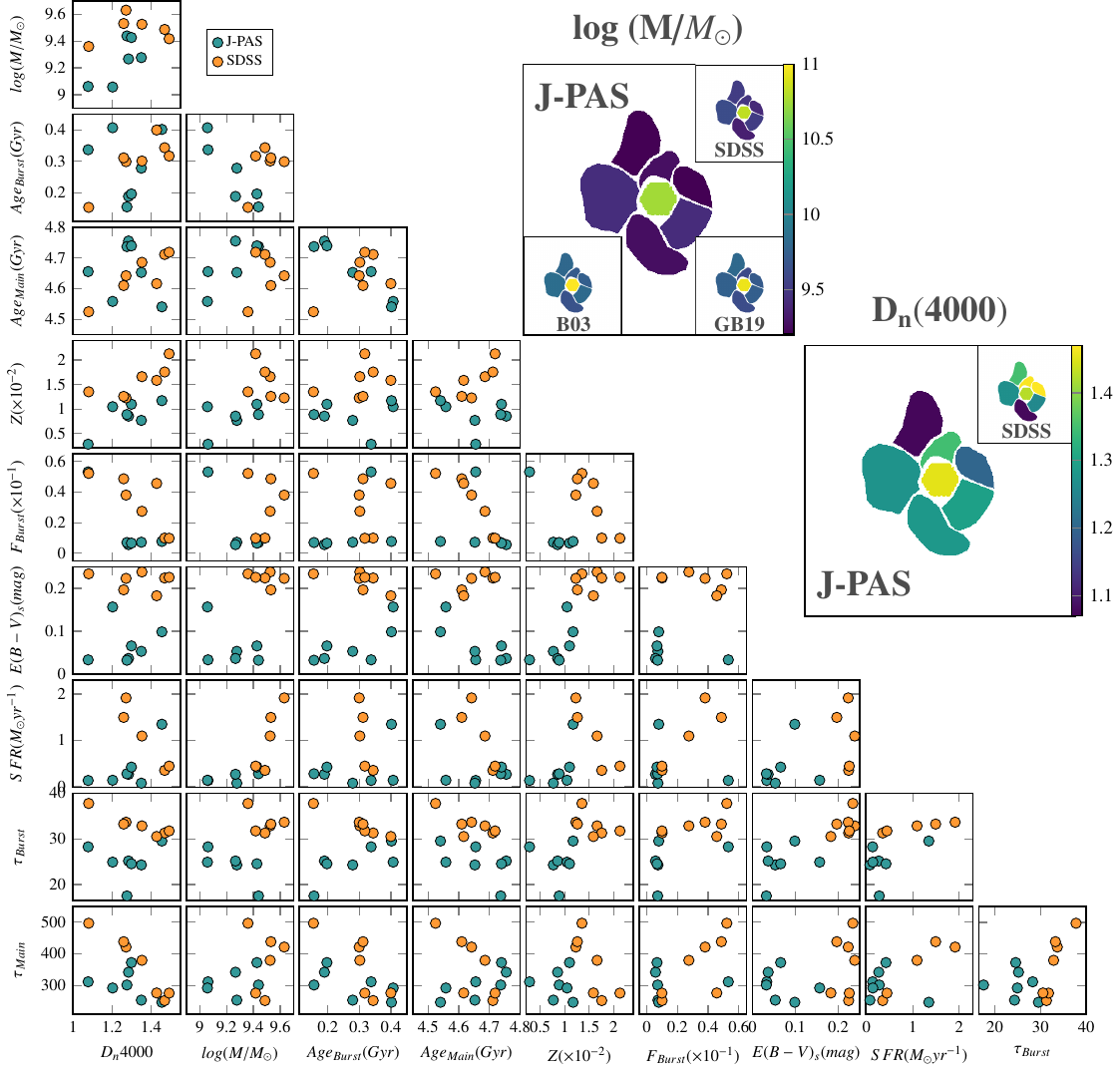}
  \fi

  \end{center}
  \caption{\label{fig:cornerplot}
    Corner plot showing the pairwise comparison of SED fitting output properties for J-PAS (teal circles) and SDSS (orange circles).
    In this analysis, all starburst and nebular components are included, and the \citet{chabrier03} IMF is applied.
    Details of the grid parameters are provided in Appendix~\ref{appendix:cigaleconfigure}.
    The two-dimensional plots in the top right illustrate the comparison between J-PAS and SDSS in terms of the $D_n(4000)$ index and stellar mass distribution.
    In addition, the stellar mass derived from SED fitting (from J-PAS and SDSS) is compared using two different methods: \citet[][is labeled as B03]{bell2003} and \citet[][is labeled as GB19]{benito19}, both based on the M/L ratio.
    The data used to generate this corner plot for both datasets are available on Zenodo; for further details, see Appendix~\ref{appendix:zenodocornerplots}.}
\end{figure*}

\subsection{SED fitting properties from J-PAS and SDSS}

In principle, Alba's current state should have a single narrative.
However, when analyzed using J-PAS and SDSS, different interpretations emerge.
Some properties are constrained by both datasets, others are constrained by only one, and certain parameters remain unconstrained in both cases.
Table~\ref{tbl:cigalout} summarizes the physical properties inferred from the SED fitting of both surveys, while Fig.~\ref{fig:cornerplot} provides a corner plot to visually compare these parameters.
In this plot, teal points denote the J-PAS results, while orange points correspond to those from SDSS.

Based on Table~\ref{tbl:cigalout}, both J-PAS and SDSS provide consistent constraints on $Age_{Main}$.
The derived values and their corresponding error bars are nearly identical between the two datasets, and, on average, the polygons yield an $Age_{Main}$ value of $4.65Gyr$.
In contrast, several parameters are constrained (value larger than the error) only by J-PAS, including $D_n(4000)$, $E(B-V_{s})$, and stellar mass, each with noticeably smaller error bars compared to SDSS.
This is because J-PAS is able to resolve spectral breaks with better precision, thereby extracting information that SDSS cannot capture as effectively.
Nevertheless, a number of parameters remain weakly constrained in both surveys, as reflected in their large uncertainties.
These include $Age_{Burst}$, metallicity, $F_{Burst}$, $\tau_{Burst}$, and $\tau_{Main}$.
This is most likely because Alba does not display any significant starburst features in its SED.
As a result, these parameters lack physical reliability in this case and cannot be incorporated into our interpretation with confidence.
We assume that if Alba was in an earlier phase of its merger (before the more massive stars had died out) these would be better constrained.
This hypothesis will be tested in followup studies on a larger sample of mergers.

The main differences observed in the corner plot of Fig.~\ref{fig:cornerplot} involve metallicity, attenuation, and stellar mass, all of which are lower in J-PAS compared to SDSS.
This result is consistent with the redder SED observed in SDSS and may reflect systematic differences in the derivation of these parameters.
However, trends such as $Age_{Burst}$ vs. attenuation, $Age_{Main}$ vs. Z, and Z vs. attenuation are likely sensitive to degeneracies.
In addition, J-PAS produces smaller values for the star formation rate (SFR), $\tau_{Burst}$, and $\tau_{Main}$.
For example, SDSS predicts a more extended burst history (both in the $\tau_{Burst}$ and in the $F_{Burst}$) leading to higher inferred SFR values.
Nevertheless, these interpretations should be approached with caution at this stage.
As noted previously, the error bars associated with $Age_{Burst}$, $\tau_{Burst}$, and $\tau_{Main}$ are large in both datasets, limiting their reliability for robust physical conclusions, even though the general correlations remain discernible.

It should also be emphasized that this analysis is based on a single case study.
A larger statistical investigation of merger remnants will be necessary to confirm and refine these trends.
Overall, for Alba's tidal features, SDSS indicates a more metal-rich stellar population with stronger attenuation, highly extended SFH, and with higher SFRs.
In contrast, J-PAS suggests a systematically lower metallicity with moderate extinction and a more rapid star formation history, consistent with a stellar population that is in an earlier phase of quenching.

As mentioned earlier, several emission lines are evident in the SDSS SED fitting of two polygons shown in Fig.~\ref{fig:fitcigale}, this could imply CIGALE is putting too much significance to the starburst component for the SDSS photometry of Alba; and making it hard to compare SDSS stellar populations with J-PAS.
This serves as a useful check to determine whether the main difference in the fraction of the starburst component derived from the SDSS data arises from degeneracies in the fits, and whether the larger fraction of the young stellar component leads to meaningful variations in $E(B-V_{s})$ and metallicity.
To verify the effect of the starburst component on the comparison, we performed an additional analysis by re-running CIGALE, this time excluding the starburst and nebular components.
Further methodological details are provided in Appendix~\ref{appendix:nostarburst}, and the corresponding results are presented in Fig.~\ref{fig:cornerplotnosfr}.
When looking at panel of $E(B-V_{s})$ vs. $Z$ in Figs.~\ref{fig:cornerplot} and \ref{fig:cornerplotnosfr}), we can confirm that the SDSS values do not change significantly.
This can also be verified by the respective error bars in Table~\ref{tbl:cigalout} (for example the error bar in SDSS's SFR is almost three times larger than the value).
This confirms that the emission lines in SDSS fittings are due to degeneracies in the fitting process and have no statistical significance on our analysis.

For the remaining parameters, the variations are minimal.
This result is consistent with the fact that Alba exhibits neither significant star formation activity nor strong emission lines.
Consequently, even after the exclusion of the starburst components, the changes properties constrained by J-PAS and SDSS is small and we can not see big difference in the constrained properties.

\subsection{Self-consistency}
\label{sec:selfconsistency}

In the previous section, we compared J-PAS and SDSS stellar population properties with each other.
In this section, we will compare the results of each with itself: seeing which properties in each filter set can be derived sufficiently accurately in the case of Alba for further study.
We therefore attempt to exclude properties that do not exhibit statistically significant variation across the various polygons.
In other words, we focus on identifying those properties with relatively small error margins in comparison to their variation across the six polygons of Alba.

To do this, we define the ``property variation significance'' (PVS) which is calculated using the standard deviation of each property (e.g., mass, age, etc.) across the six polygons and dividing it by the median of that property's error bar (also across the six polygons).
The formula used is:

\begin{equation}
  \label{eq:criterion}
  {PVS_{property}=\frac{STD(property\_in\_six\_polygons)}{Median(error\_of\_property\_in\_six\_polygons)}}
\end{equation}

The PVS therefore quantifies the significance of the scatter of a property in the various components relative to the representative error of that property.
For instance, in the case of stellar mass, the numerator of the equation for J-PAS is {$\massjpasnum\times10^8M_\odot$} while the denominator is {$\massjpasden\times10^8M_\odot$} (note that, since we are considering the ratio, the logarithmic mass is not used).
This yields a ratio of $2.38\sigma$, indicating that the scatter in this property among the polygons is comparatively high compared to their noise/error.
In contrast, for the same property measured using SDSS, the values are {$\masssdssnum\times10^8M_\odot$} and {$\masssdssden\times10^8M_\odot$}, producing a ratio of $0.70\sigma$.
This result suggests that, for SDSS, the variation in stellar mass values across Alba's tidal features is less significant (harder to interpret) in comparison to J-PAS.

We rank the output properties studied here according to their PVS in Table~\ref{tbl:criterion} (sorted by their PVS in J-PAS).
For the rest of this paper, properties with $\rm{PVS}<1.5$ are excluded, as their variation across polygons is within the margin of error (less than $1.5\sigma$), rendering their differences unreliable for analysis of the Alba system with the current data and CIGALE run.

Alba is in the later stage of a major merger, hence the stellar populations of the progenitors are heavily mixed and the most prominent/massive OB stars of a potential burst in star formation (triggered by the merger) have died out.
Therefore, it is reasonable that the PVS of some properties like age of burst is not distinguishable (within the error bars) even with J-PAS resolution in Alba.
For example, in earlier phases of a non-dry merger (when the OB stars from a possibly induced star formation are still present), J-PAS resolution should be able to determine the age of burst with much smaller errors.
Follow-up studies on a larger sample of mergers in J-PAS will be done in our next papers to assess this.

\begin{table}[t]
  \caption{The property variation significance (PVS) of the CIGALE outputs presented in Table~\ref{tbl:cigalout} is evaluated for both J-PAS and SDSS.
  The dataset used to produce this table is publicly available on Zenodo, further details can be found in Appendix~\ref{appendix:zenodotabels}.}
\label{tbl:criterion}
\small
\centering
\begin{tabular}{lcc}
\hline\hline
Property &
PVS (J-PAS) &
PVS (SDSS) \\
\hline
log (M/\(M_\odot\))               &    $2.38$    &    $0.70$ \\
$D_n(4000)$                       &    $1.63$    &    $0.72$ \\
$Age_{Burst}$                     &    $0.48$    &    $0.08$ \\
$\tau_{Main}$                     &    $0.34$    &    $0.18$ \\
SFR                               &    $0.31$    &    $0.25$ \\
$E(B-V)_{s}$                      &    $0.29$    &    $0.12$ \\
$Age_{Main}$                      &    $0.12$    &    $0.10$ \\
$\tau_{Burst}$                    &    $0.12$    &    $0.23$ \\
Fraction Burst                    &    $0.04$    &    $0.25$ \\
Z                                 &    $0.01$    &    $0.06$ \\ \hline

\end{tabular}
\end{table}

The $M_\star$ and $D_n(4000)$ from the J-PAS dataset met the $\rm{PVS}>1.5$ condition, with values of $2.38$ and $1.63$, respectively.
In contrast, none of the output properties derived from SDSS photometry passed this condition.
Therefore, for Alba (this case-study), our analysis focuses on stellar mass and $D_n(4000)$.
Depending on the physical characteristics of different mergers, we employ the PVS framework to identify the properties most affected at the observed phase.
In this sense, PVS functions as a calibrator of galaxy mergers, highlighting the properties whose variations are most informative within the constraints of the available data for each galaxy.
This approach enables the classification of merging systems according to their distinctive stellar population or gas/nebula features, a task that was previously unattainable.
For example, once extended to many merging galaxies, we can extract galaxies which exhibit PVS($D_n(4000)$) greater than 1.5 to examine potential correlations with merger dynamics, morphology, or environment.

After evaluating the self-consistency of J-PAS and SDSS and finding the parameters that can yield useful information for Alba, we can now look into those two parameters.
The $D_n(4000)$ index quantifies the strength of the 4000 {\AA} break in the optical spectrum of a galaxy \citep{balogh99}, which is primarily influenced by the age and metallicity of its stellar population \citep{kauffmann03, gallazzi05}.
As demonstrated by \citet{kauffmann03}, the $D_n(4000)$ index increases as a function of stellar population age.
In Fig.~\ref{fig:cornerplot} (bottom-right panel), we see the $D_n(4000)$ values across the 2D spatial footprint of the Alba galaxy.
In J-PAS this spatial distribution shows that the center has the highest value.
This is expected in a major merger: the fraction of younger stars in the center is the least since any possible gas in the progenitors was stripped or turned into stars in earlier phases.
Also, the North-1 has the smallest value that is consistent with its bluer color in the HSC image of Figure \ref{fig:colorimg}.
But in SDSS, we see that the center doesn't have the largest value and the spatial distribution is hard to interpret.

For Alba, the dominant factor influencing $D_n(4000)$ appears to be the recent major merger event and the resulting star formation (which has become quiescent ever since), but nevertheless different in each of the polygons (that contain stars from different phases of the merger).
Further simulations could help constrain essential merger parameters, such as the mass ratio of the interacting galaxies, the impact parameter, and the time since the merger occurred, but such investigations are beyond the scope of this paper.

The stellar mass ($M_\star$) also passed the $\rm{PVS}>1.5$ condition for Alba with J-PAS.
In Fig.~\ref{fig:cornerplot} (top-right panel), we present the stellar mass distribution across different regions of the polygons in a 2D map of the Alba galaxy.
The J-PAS data show that the stellar mass in the tidal streams is systematically lower (darker colors) than that derived from the SDSS data and with less scatter.
Except for these two issues, the overall 2D mass distribution is consistent between J-PAS and SDSS, with the central region being the most massive (with approximately similar masses).

However, calculating the ratio between the diffuse components and the central region to estimate the merger mass ratio is not feasible in this case.
Alba is already in a late stage of a major merger, where the central regions of the progenitor systems have coalesced.
Consequently, the mass ratio, which is typically determined before coalescence, cannot be reliably derived here.

\subsection{Stellar mass from SED fitting and heuristic methods}

SED fitting is not the only method employed in the literature to estimate stellar mass from SDSS filters; there are also heuristic methods to only derive the stellar mass, for example B03 and GB19.
Therefore it is necessary to also evaluate our results with those.
B03 has been the most widely used for stellar streams \citep[e.g.,][among many others]{martinezb23,elizabeth25}.
However, B03 is based on a \citet{salpeter55} IMF, while our default run of CIGALE uses a \citet{chabrier03} IMF.
We have approached this issue in two ways: 1) we used the GB19 heuristic which also assumes a Chabrier IMF. 2) we re-run CIGALE with a Salpeter IMF (see Appendix~\ref{appendix:salpeter}).

To derive a heuristic estimate of the stellar mass, we calculate the absolute magnitude of the object in a selected SDSS filter, such as the r-band ($M_{r}$).
As an example, we illustrate the process using the r-band; however, the same procedure is applied across all SDSS filters.
This calculation incorporates the apparent magnitude, distance modulus, and extinction.
We obtain the absolute magnitude of the Sun in the r-band $M_{\odot,r}$, from \citet{willmer18}.
Using the luminosity–magnitude relation we calculate:
\begin{equation}
  \label{eq:absmag}
  {M=-2.5\times log_{10}(L_r)}
\end{equation}
the luminosities of both the object and the Sun in the r-band.
We then rearrange this equation to yield the luminosity in units of solar luminosity:
\begin{equation}
  \label{eq:luminosity}
  {\frac{L_{r}}{L_{\odot}}=10^{0.4\times(M_{\odot}-M_{r})}}
\end{equation}
We calculated the stellar mass by multiplying the luminosity ratio by the mass-to-light ratio in the r-band, using the coefficients provided by B03 (Table A.7 in their paper) and GB19 (Table B.1 in their paper).
  Finally, we obtain the $g-r$, $g-z$, and $r-z$ colors, and derive the stellar mass for each filter/color combination (yielding 9 masses for the three colors).
  We use the mean of these colors to determine the mass-to-light ratio and their standard deviation to define the scatter of the distribution, following the methods of \citet{martinezb23} and \citet{elizabeth25}.
To validate our procedure, we tested our scripts to the magnitudes and colors of \citet{martinezb23} and successfully reproduced their stellar mass estimates.
Fig.~\ref{fig:cornerplot} presents the results of the stellar mass estimates derived from B03 and GB19 for Alba’s polygons. These are visualized as 2D maps in the stellar mass panel, with the lower-left corner showing the B03 results and the lower-right corner displaying those from GB19.

The color bar in the plot was adjusted to account for all four displayed results; therefore, the difference between the B03 and GB19 models is not immediately visible.
To compare the two methods quantitatively, we converted the stellar masses to a linear scale and found that the B03 estimates are approximately 1.2 times higher than those of GB19.
This outcome is expected, as the \citet{salpeter55} IMF is bottom-heavy compared to the \citet{chabrier03} IMF, naturally resulting in higher stellar mass estimates for B03.

Our default run of CIGALE employed a \citet{chabrier03} IMF; therefore, we re-ran the code using a \citet{salpeter55} IMF to evaluate its effect on the stellar mass derived for Alba’s streams with J-PAS and SDSS SED fitting (see Fig.~\ref{fig:cornerplotsalpiter} in Appendix~\ref{appendix:salpeter}).
The analysis shows that adopting the \citet{salpeter55} IMF produces stellar masses 1.73 and 1.76 times larger than those obtained with the \citet{chabrier03} IMF for J-PAS and SDSS, respectively.
This result is consistent across both datasets, as the bottom-heavy \citet{salpeter55} IMF naturally yields higher stellar mass estimates than the \citet{chabrier03} IMF.

Comparing the heuristic stellar masses with the SED-based stellar masses (J-PAS or SDSS), we see that the heuristic methods produce systematically higher stellar masses than those derived from SED fitting.
When comparing to the SED fitting results for Alba's tidal features, the difference is mostly larger than 0.4 dex ($\sim2.51$) for SDSS and 0.5 dex ($\sim3.16$) for J-PAS.

\section{Discussion}
\label{sec:discussion}
After deriving the relevant properties from both the central region and the tidal features outlined by the polygons, we can now assess our understanding of the physical processes in Alba.

To confirm that Alba represents an isolated merger event, we examined all nearby objects listed in the DESI catalog within a radial redshift range of ±0.5 Mpc.
The nearest galaxy located outside the polygons in Fig.~\ref{fig:colorimg} lies at a tangential distance of 1 Mpc, supporting the conclusion that the merger remnants in Alba are gravitationally bound and that none of the progenitor components has escaped.
Furthermore, the bright object visible in the northwest region of Fig.~\ref{fig:colorimg}, at coordinates 246.749208 and 43.486972 deg, is identified as a foreground star.
Its large separation ensures that it does not affect the system under study and, therefore, does not require PSF subtraction at the depth of J-PAS/SDSS.

The $D_n(4000)$ index is a well-established tracer of the state of star formation; for instance, \citet{zahid17} adopt a threshold of $D_n(4000)$ $\geq1.5$ to define a quiescent population.
In the 2D $D_n(4000)$ map from J-PAS Fig.~\ref{fig:cornerplot}, polygon West-1 displays the second lowest $D_n(4000)$ value after North-1, while the remaining polygons exhibit similar values.
Compared with North-1, West-1 has a comparable stellar mass and star formation rate but shows the highest attenuation.

In the SDSS data, the higher $D_n(4000)$ values of West-1 and North-2 relative to the central polygon cannot be reliably interpreted, as these two regions have the largest error bars among all polygons.
Consequently, any apparent differences with the center should be treated with caution.

Applying the $D_n(4000)$ analysis to Alba’s merger remnants suggests that they have not yet fully reached the quiescent phase.
This interpretation is intriguing, as any gas present in the progenitor galaxies would have been expected to form stars during the earlier stages of the merger, prior to the coalescence at the center of mass.

Overall, for all polygons, the corner plot in Fig.~\ref{fig:cornerplot} shows that metallicity, attenuation, and stellar mass are lower in J-PAS compared to SDSS.
This observation aligns with the redder SEDs seen in SDSS and may indicate systematic differences in the methods used to derive these parameters.
Nevertheless, relationships such as $Age_{Burst}$ versus attenuation, $Age_{Main}$ versus metallicity, and metallicity versus attenuation are likely influenced by degeneracies.

Additionally, J-PAS yields smaller values for the star formation rate (SFR), $\tau_{Burst}$, and $\tau_{Main}$.
Therefore SDSS suggests a more extended burst history, both in the main stellar age and the recent burst, resulting in higher inferred SFRs.
Interestingly, $Age_{Burst}$, and $Age_{Main}$ remain broadly consistent between the two surveys.
However, these results should be interpreted with caution, as the associated error bars for $Age_{Burst}$, $\tau_{Burst}$, and $\tau_{Main}$ are large in both datasets, limiting their reliability for drawing robust physical conclusions, even though the overall trends are still apparent.

\citet{fran25} reported no significant discrepancies between stellar masses derived from CIGALE SED fitting and those estimated using M/L ratios.
However, their study differs from ours in several key aspects.
They employed J-PLUS data, which includes 12 filters (five SDSS broad bands and seven medium/narrow bands), whereas we use J-PAS data with 54 narrow-band and two medium-band filters.
Moreover, their analysis was performed statistically across a sample of galaxies using integrated fluxes (one value per galaxy), while our study focuses on individual tidal features of a single galaxy undergoing a major merger.

Through the PVS, we find properties that have the most statistically meaningful variation across the Alba merging system, compared to their errors.
However, looking at Table.~\ref{tbl:criterion} another systematic trend also becomes visible between SDSS and J-PAS for some of the properties where the $PVS<1.5$: metallicity, Fraction burst, $E(B-V)_{s}$ and SFR.
Comparing the J-PAS values to SDSS values directly, we see that they are systematically much lower than SDSS (also in the errors).
For example, the median J-PAS fraction of burst among the tidal features (not including the center) is $0.07$ with a median error of $0.17$ while for SDSS, it is $0.35$ with a median error of $0.69$.
So while their PVS is low in both SDSS and J-PAS, the absolute values when comparing SDSS to J-PAS are meaningfully distinguishable: the J-PAS value falls within the error of SDSS (is only $0.4\sigma$ away from the median SDSS), but not the inverse (the SDSS value is $1.6\sigma$ away from the median J-PAS).
In followup studies on larger sample of galaxies, such improvements will enable better comparisons of the over-all merger population (not just their components).

All of the interpretations presented here are based on a single case study.
To ensure the robustness of these results, a statistical analysis of merger remnants is required.
The J-PAS survey provides an opportunity to carry out such analyses on a large scale, enabling the blind selection of this type of galaxies.
In addition, simulations such as those by \citet[see in particular their fig.~1]{petersson23} are essential for tracing the evolutionary history of galaxies and for establishing a meaningful comparison between observations and theoretical models.

\section{Summary and conclusions}
\label{sec:conclusions}

In this pilot study, we investigate the complex structures of an ongoing major merging galaxy (PGC 3087775, or Alba for short) to determine which stellar population properties of its tidal features benefit the most from the high spectral resolution provided by the J-PAS filter system and use them to get some insight into this particular merger.
Alba is the most prominent/largest/brightest tidal feature observed in the 12 $deg^{2}$ of the J-PAS Early Data Release (EDR).
We compare SED fitting results of J-PAS with those obtained using the SDSS filter set, which has traditionally dominated studies of such structures.
To our knowledge, this is the first time that the spectral energy distribution (SED) of a tidal feature has been contiguously mapped along its entire length at this level of spectral detail.
The main results of this study are summarized as follows:

\begin{itemize}
\item J-PAS reaches a \emph{continuum} $3\sigma$ measured surface brightness limit (mSBL, over 100 arcsec$^2$) of 26.57 mag/arcsec$^2$ or $5.2\times10^{-20}$ erg/s/cm$^2$/{\AA}/arcsec$^2$ at 7000~{\AA}.
  \item The mSBL in the red filters is higher compared to the bluer filters, primarily due to the presence of telluric emission lines, which introduce increased Poisson noise and degrade the data quality in affected wavelengths.
  \item Comparison with MaNGA demonstrates the excellent calibration of J-PAS (see Fig.~\ref{fig:manga}).
  \item All identified merger remnants in Alba exhibit a clear $D_n(4000)$ break, consistent with the presence of intermediate age stellar populations.
    The average $D_n(4000)$ index of the tidal features is 1.24, suggesting that the merger was not a fully dry merger: some star formation has occurred during the merger that has not yet fully quenched.
  \item Both J-PAS and SDSS can constrain the $Age_{Main}$, whereas $D_n(4000)$, dust attenuation, and stellar mass are constrained exclusively by J-PAS.
    The remaining properties, however, remain unconstrained in both surveys.
  \item SDSS predicts a metal-rich and highly extended star formation history (SFH) characterized by elevated star formation rates.
    In contrast, J-PAS suggests a less metal-rich population with moderate extinction and a more rapid SFH, consistent with an already quenched stellar population.
  \item We excluded the starburst and nebular components and found that the CIGALE output parameters related to the stellar population (excluding gas and emission lines) remained largely unaffected in SDSS.
    This is due to degeneracies with the SDSS filters in Alba's tidal features.
  \item Even for a small PVS, properties derived from J-PAS are significantly more precise in comparison with SDSS.
  \item Stellar mass estimates derived from SDSS photometry reveal that the values reported by B03 are approximately 1.2 times higher than those of GB19, a result consistent with theoretical expectations, since the \citet{salpeter55} IMF is more bottom-heavy than the \citet{chabrier03} IMF and therefore naturally yields higher stellar mass estimates in B03.
  \item When applied to SED fitting, the \citet{salpeter55} IMF yields stellar mass estimates that are 1.73 times higher for J-PAS and 1.76 times higher for SDSS compared to those derived with the \citet{chabrier03} IMF; this systematic difference reflects the bottom-heavy nature of the \citet{salpeter55} IMF, which inherently produces larger stellar mass values across both datasets.
  \item Relative to SED-based estimates from J-PAS and SDSS, heuristic methods systematically yield higher stellar masses, with discrepancies exceeding 0.4 dex ($\sim2.51$) for SDSS and 0.5 dex ($\sim3.16$) for J-PAS when applied to Alba’s tidal features.
\end{itemize}

J-PAS provides, for the first time, the ability to obtain IFU-like SEDs over wide sky areas that have remained inaccessible at this spectral resolution.
J-PAS therefore can detect much more subtle differences in the stellar populations of the faint remnants of the mergers.
This will be used in future studies on a larger population of merging galaxies at various stages; shedding new light on our understanding of galaxy evolution and the role of mergers in stellar mass assembly.

\section{Code and data availability}

This project was developed in the reproducible framework of Maneage \citep[\emph{Man}aging data lin\emph{eage},][latest Maneage commit \texttt{\maneageversion}, from \maneagedate]{maneage}.
It was built on an {\machinearchitecture} machine with {\machinebyteorder} byte-order, see Appendix \ref{appendix:software} for the used software and their versions.
The source code for the project is publicly available on SoftwareHeritage for longevity as \href{https://archive.softwareheritage.org/swh:1:dir:7c203a565ec6ef40d8419dbdf70e8639b78f4b99;origin=https://gitlab.com/Sepideh.Esk/alba;visit=swh:1:snp:adc584ff3540ea3151de6ebc8462805e8818b1f4;anchor=swh:1:rev:f9d0f9cb25c80415107c962cc53d369a650e9b82}{\texttt{swh:1:dir:7c203a56\-5ec6ef40\-d8419dbdf\-70e8639b7\-8f4b99}}\footnote{SoftWare Hash IDentifier (SWHID) can be used with resolvers such as \texttt{http://n2t.net/} (e.g., \texttt{http://n2t.net/swh:1:...}). Clicking on the SWHIDs will provide more ``context'' for same content.} and all the output data products are available on \href{https://doi.org/10.5281/zenodo.17674805}{zenodo.17674805}\footnote{\url{https://doi.org/10.5281/zenodo.17674805}}.
For the underlying data of the figures and tables see Appendix~\ref{appendix:zenododataforplots}.

\vspace{5mm}

\begin{acknowledgements}
  We would like to extend our sincere gratitude to Andr\'es del Pino, Borja Anguiano, Alvaro Alvarez-Candal,   Jos\'e Mar\'ia Diego Rodr\'igiez, and Eduardo Telles.

  This paper has gone through internal review by the J-PAS collaboration.
  The authors acknowledge the following grants by the Spanish Ministry of Science and Innovation (MCIN\-/AEI/\-10.13039/\-501100011033 y FEDER, Una manera de hacer Europa): PID\-2021-1249\-18NB-C41, PID\-2021-12491\-8NB-C42, PID\-2021-1249\-18NA-C43, PID\-2021-1249\-18NB-C44, PID\-2022-1365\-05NB-I00, PID\-2022-1417\-55NB-I00, PID\-2024-16222\-9NB-I00 and Severo Ochoa grant CEX2021-001131-S as well as European Union MSCA Doctoral Network EDUCADO, GA 101119830 and Widening Participation, ExGal-Twin, GA 101158446.

  This work is based on observations made with the JST250 telescope and JPCam camera for the J-PAS project at the Observatorio Astrof\'{\i}sico de Javalambre (OAJ), in Teruel, owned, managed, and operated by the Centro de Estudios de F\'{\i}sica del  Cosmos de Arag\'on (CEFCA). We acknowledge the OAJ Data Processing and Archiving Department (DPAD) for reducing and calibrating the OAJ data used in this work.
Funding for the J-PAS Project has been provided by the Governments of Spain and Aragón through the Fondo de Inversión de Teruel, European FEDER funding and the Spanish Ministry of Science, Innovation and Universities, and by the Brazilian agencies FINEP, FAPESP, FAPERJ and by the National Observatory of Brazil. Additional funding was also provided by the Tartu Observatory and by the J-PAS Chinese Astronomical Consortium.

Funding for the Sloan Digital Sky Survey IV has been provided by the Alfred P. Sloan Foundation, the U.S. Department of Energy Office of Science, and the Participating Institutions.
SDSS-IV acknowledges support and resources from the Center for High Performance Computing  at the University of Utah. The SDSS website is www.sdss4.org.
SDSS-IV is managed by the Astrophysical Research Consortium for the Participating Institutions of the SDSS Collaboration including the Brazilian Participation Group, the Carnegie Institution for Science, Carnegie Mellon University, Center for Astrophysics | Harvard \& Smithsonian, the Chilean Participation Group, the French Participation Group, Instituto de Astrof\'isica de Canarias, The Johns Hopkins University, Kavli Institute for the Physics and Mathematics of the Universe (IPMU) / University of Tokyo, the Korean Participation Group, Lawrence Berkeley National Laboratory, Leibniz Institut f\"ur Astrophysik Potsdam (AIP),  Max-Planck-Institut f\"ur Astronomie (MPIA Heidelberg), Max-Planck-Institut f\"ur Astrophysik (MPA Garching), Max-Planck-Institut f\"ur Extraterrestrische Physik (MPE), National Astronomical Observatories of China, New Mexico State University, New York University, University of Notre Dame, Observat\'ario Nacional / MCTI, The Ohio State University, Pennsylvania State University, Shanghai Astronomical Observatory, United Kingdom Participation Group, Universidad Nacional Aut\'onoma de M\'exico, University of Arizona, University of Colorado Boulder, University of Oxford, University of Portsmouth, University of Utah, University of Virginia, University of Washington, University of Wisconsin, Vanderbilt University, and Yale University.

\end{acknowledgements}

\bibliographystyle{aa}
\bibliography{references}

\begin{appendix}

\section{residual}
\label{appendix:residual}
\begin{figure*}[!t]
  \begin{center}
  \ifdefined\makepdf%
    \tikzsetnextfilename{fig-residual}%
\begin{tikzpicture}


  \begin{axis}
    [ at={(-0.74\linewidth,0.012\linewidth)},
      xmin=3400,
      xmax=9400,
      ymin=-25,
      ymax=25,
      width=0.55\linewidth,
      height=0.2\linewidth,
      enlarge x limits=false,
      enlarge y limits=false,
      xticklabel=\empty,
      legend pos=north west,
      legend columns=4,
      legend style={at={(0.5,1.25)},anchor=north},
      legend cell align={center},
      legend style={nodes={scale=0.71, transform shape}},
      yticklabel style={font=\bfseries},
    ]

    \addplot [only marks, fill=violet!90!black, mark size=2, error bars/.cd, y dir=both, y explicit]
    table [x index=0, y index=1, y error index=2] {tex/build/figures/fit-cgale-jpas/north1-residual-jpas.txt};
    \addlegendentry{Observed (J-PAS)}

    \addplot [only marks, fill=gray!85!black, mark size=3, mark=triangle*, error bars/.cd, y explicit, y dir=both]
    table [x index=0, y index=1, y error index=2] {tex/build/figures/fit-cgale-sdss/north1-residual-sdss.txt};
    \addlegendentry{Observed (SDSS)}

  \end{axis}

  \node[anchor=south] at (-0.33\linewidth,0.01\linewidth)
       {\textcolor{violet!90!black}{\bf \Large North-1}};


  \begin{axis}
    [ at={(-0.74\linewidth,-0.11\linewidth)},
      xmin=3400,
      xmax=9400,
      ymin=-25,
      ymax=25,
      width=0.55\linewidth,
      height=0.2\linewidth,
      enlarge x limits=false,
      enlarge y limits=false,
      xticklabel=\empty,
      yticklabel=\empty,
    ]

    \addplot [only marks, fill=green!90!black, mark size=2, error bars/.cd, y dir=both, y explicit]
    table [x index=0, y index=1, y error index=2] {tex/build/figures/fit-cgale-jpas/east-residual-jpas.txt};

    \addplot [only marks, fill=gray!85!black, mark size=3, mark=triangle*, error bars/.cd, y explicit, y dir=both]
    table [x index=0, y index=1, y error index=2] {tex/build/figures/fit-cgale-sdss/east-residual-sdss.txt};

  \end{axis}

  \node[anchor=south] at (-0.33\linewidth,-0.11\linewidth)
       {\textcolor{green!90!black}{\bf \Large East}};

  \node[anchor=south, rotate=90, scale=0.75] at (-0.75\linewidth,-0.05\linewidth)
       {\textcolor{black!90!black}{$\huge\bf \Delta\%$}};


  \begin{axis}
    [ at={(-0.74\linewidth,-0.232\linewidth)},
      xmin=3400,
      xmax=9400,
      ymin=-25,
      ymax=25,
      width=0.55\linewidth,
      height=0.2\linewidth,
      enlarge x limits=false,
      enlarge y limits=false,
      yticklabel style={font=\bfseries},
      xlabel={Observed Wavelength [{\AA}]},
    ]

    \addplot [only marks, fill=blue!80!white, mark size=2, error bars/.cd, y dir=both, y explicit]
    table [x index=0, y index=1, y error index=2] {tex/build/figures/fit-cgale-jpas/south-residual-jpas.txt};

    \addplot [only marks, fill=gray!85!black, mark size=3, mark=triangle*, error bars/.cd, y explicit, y dir=both]
    table [x index=0, y index=1, y error index=2] {tex/build/figures/fit-cgale-sdss/south-residual-sdss.txt};

  \end{axis}

  \node[anchor=south] at (-0.33\linewidth,-0.23\linewidth)
       {\textcolor{blue!90!black}{\bf \Large South}};


  \begin{axis}
    [ at={(-0.26\linewidth,0.012\linewidth)},
      xmin=3400,
      xmax=9400,
      ymin=-25,
      ymax=25,
      width=0.55\linewidth,
      height=0.2\linewidth,
      enlarge x limits=false,
      enlarge y limits=false,
      xticklabel=\empty,
      yticklabel=\empty,
      legend pos=north west,
      legend columns=4,
      legend style={at={(0.5,1.25)},anchor=north},
      legend cell align={center},
      legend style={nodes={scale=0.71, transform shape}},
    ]

    \addplot [only marks, fill=violet!60!white, mark size=2, error bars/.cd, y dir=both, y explicit]
    table [x index=0, y index=1, y error index=2] {tex/build/figures/fit-cgale-jpas/north2-residual-jpas.txt};
    \addlegendentry{Observed (J-PAS)}

    \addplot [only marks, fill=gray!85!black, mark size=3, mark=triangle*, error bars/.cd, y explicit, y dir=both]
    table [x index=0, y index=1, y error index=2] {tex/build/figures/fit-cgale-sdss/north2-residual-sdss.txt};
    \addlegendentry{Observed (SDSS)}

  \end{axis}

  \node[anchor=south] at (0.15\linewidth,0.01\linewidth)
       {\textcolor{violet!60!white}{\bf \Large North-2}};


  \begin{axis}
    [ at={(-0.26\linewidth,-0.11\linewidth)},
      xmin=3400,
      xmax=9400,
      ymin=-25,
      ymax=25,
      width=0.55\linewidth,
      height=0.2\linewidth,
      enlarge x limits=false,
      enlarge y limits=false,
      xticklabel=\empty,
      yticklabel=\empty,
    ]

    \addplot [only marks, fill=red!80!black, mark size=2, error bars/.cd, y dir=both, y explicit]
    table [x index=0, y index=1, y error index=2] {tex/build/figures/fit-cgale-jpas/west1-residual-jpas.txt};

    \addplot [only marks, fill=gray!85!black, mark size=3, mark=triangle*, error bars/.cd, y explicit, y dir=both]
    table [x index=0, y index=1, y error index=2] {tex/build/figures/fit-cgale-sdss/west1-residual-sdss.txt};

  \end{axis}

  \node[anchor=south] at (0.15\linewidth,-0.11\linewidth)
       {\textcolor{red!80!black}{\bf \Large West-1}};


  \begin{axis}
    [ at={(-0.26\linewidth,-0.232\linewidth)},
      xmin=3400,
      xmax=9400,
      ymin=-25,
      ymax=25,
      width=0.55\linewidth,
      height=0.2\linewidth,
      enlarge x limits=false,
      enlarge y limits=false,
      yticklabel=\empty,
      xlabel={Observed Wavelength [{\AA}]},
    ]

    \addplot [only marks, fill=red!60!white, mark size=2, error bars/.cd, y dir=both, y explicit]
    table [x index=0, y index=1, y error index=2] {tex/build/figures/fit-cgale-jpas/west2-residual-jpas.txt};

    \addplot [only marks, fill=gray!85!black, mark size=3, mark=triangle*, error bars/.cd, y explicit, y dir=both]
    table [x index=0, y index=1, y error index=2] {tex/build/figures/fit-cgale-sdss/west2-residual-sdss.txt};

  \end{axis}

  \node[anchor=south] at (0.15\linewidth,-0.23\linewidth)
       {\textcolor{red!60!white}{\bf \Large West-2}};

\end{tikzpicture}%
  \else
    \includegraphics[width=\linewidth]{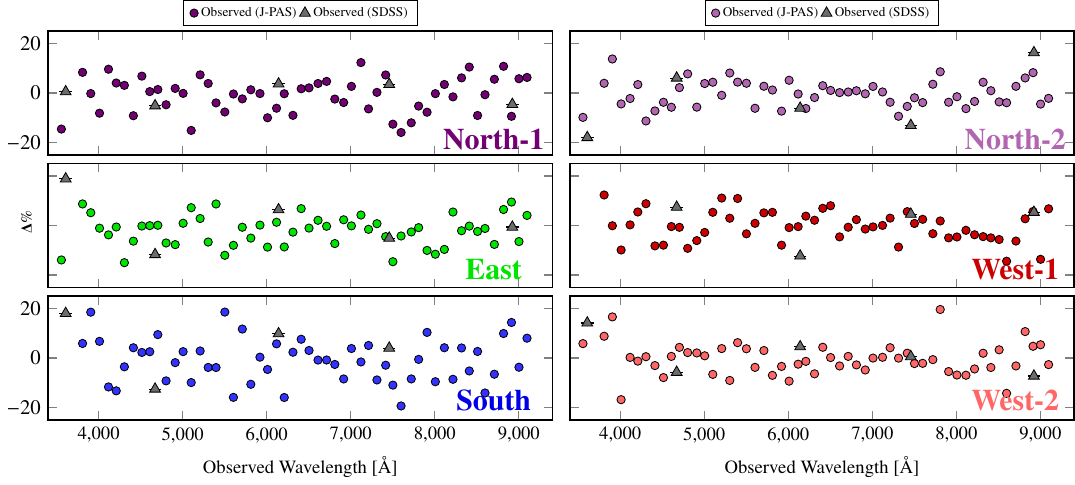}
  \fi

  \end{center}
  \caption{\label{fig:figresidual}
    Residuals between the SED fitting results from J-PAS and SDSS surveys and their corresponding observed fluxes.
    The circles represent the residuals associated with the J-PAS data, while the triangles indicate those derived from SDSS.
    The color coding of the J-PAS data matches that of the polygons shown in Fig.~\ref{fig:colorimg}, ensuring consistency across the figures and facilitating visual comparison.
  The data used to generate these residual plots are available on Zenodo; further details can be found in Appendix~\ref{appendix:zenodofigresidual}.}
\end{figure*}
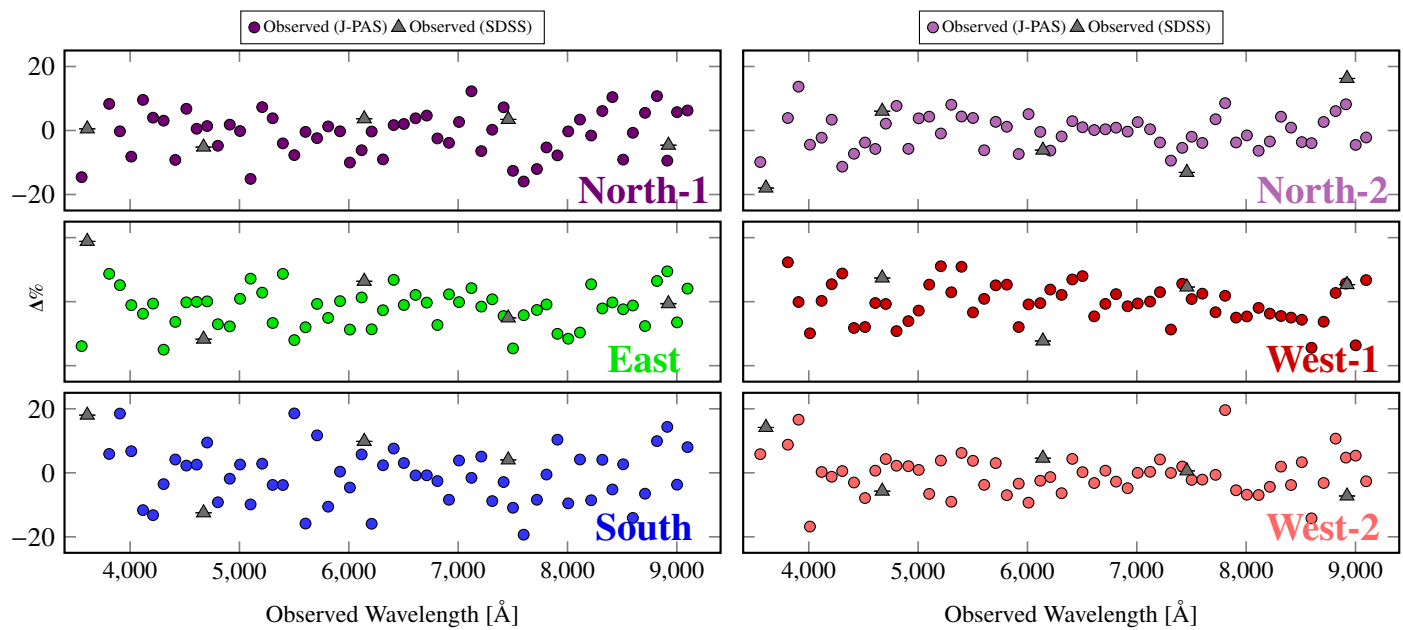

The residuals between the SED fitting results obtained from CIGALE and their corresponding observed fluxes are presented in Fig.~\ref{fig:figresidual}.
These residuals are displayed using the same color scheme adopted in Fig.~\ref{fig:fitcigale}, ensuring consistency and facilitating direct comparison between the fitted models and the observational data.
This representation allows for a clearer assessment of the quality of the SED fitting across the different regions analyzed.

\section{CIGALE configuration for SED fitting}
\label{appendix:cigaleconfigure}

The main grid parameters modified in the CIGALE configuration file are listed below.
For further details, the complete configuration file is available on Zenodo.

\begin{verbatim}
tau_main = 50.0,  100.0, 500.0, 1000.0
age_main = 500.,  1000., 2000., 3000., 4000.,
           4025., 4050., 4075., 4100., 4125.,
           4150., 4175., 4200., 4225., 4250.,
           4275., 4300., 4325., 4350., 4375.,
           4400., 4425., 4450., 4475., 4500.,
           4525., 4550., 4575., 4600., 4625.,
           4650., 4675., 4700., 4725., 4750.,
           4775., 4800., 4825., 4850., 4875.,
           4900., 4925., 4950., 4975., 5000.,
           5025., 5050., 5075., 5100., 5125.,
           5150., 5175., 5200., 5225., 5250.,
           5275., 5300., 5325., 5350., 5375.,
           5400., 5425., 5450., 5475., 5500.
tau_burst   = 1.0, 5.0, 20.0, 50.0, 100.0
age_burst   = 10., 100., 500.
f_burst     = 0.001, 0.005, 0.01, 0.2
metallicity = 0.0001, 0.0004, 0.004, 0.008,
              0.02, 0.05
E_BV_lines  = 0.0, 0.1, 0.2, 0.3, 0.5, 0.7,
              0.8
E_BV_factor = 0.44
sfr_A       = 1.0
\end{verbatim}

\section{CIGALE using the Salpeter IMF}
\label{appendix:salpeter}

To compare the stellar mass obtained from the SED fitting with B03, we ran CIGALE using the \citet{salpeter55} IMF.
This choice was necessary because B03 employed the \citet{salpeter55} IMF in their mass-to-light (M/L) relation.
It is therefore essential to check how the different IMF we used in the main CIGALE SED fitting affects our result.
For this reason, we executed CIGALE with exactly the same grid parameters described in Appendix~\ref{appendix:cigaleconfigure}, but substituting the \citet{salpeter55} IMF.
The resulting parameter values are presented in the corner plot of Fig~\ref{fig:cornerplotnosfr}.

As shown in Fig~\ref{fig:cornerplotnosfr} and \ref{fig:cornerplot}, the stellar mass derived with the \citet{salpeter55} IMF is larger than that obtained with the \citet{chabrier03} IMF.
This difference may be explained by the underlying assumptions of the two models.
Specifically, the \citet{salpeter55} IMF accounts for a greater number of low-mass stars compared to the \citet{chabrier03} IMF, which leads to systematically higher stellar mass estimates.

\begin{figure*}[!t]
  \begin{center}
  \ifdefined\makepdf%
    \tikzsetnextfilename{fig-corner-plot-salpiter}%
    \input{tex/src/fig-corner-plot-salpiter.tex}%
  \else
    \includegraphics[width=\linewidth]{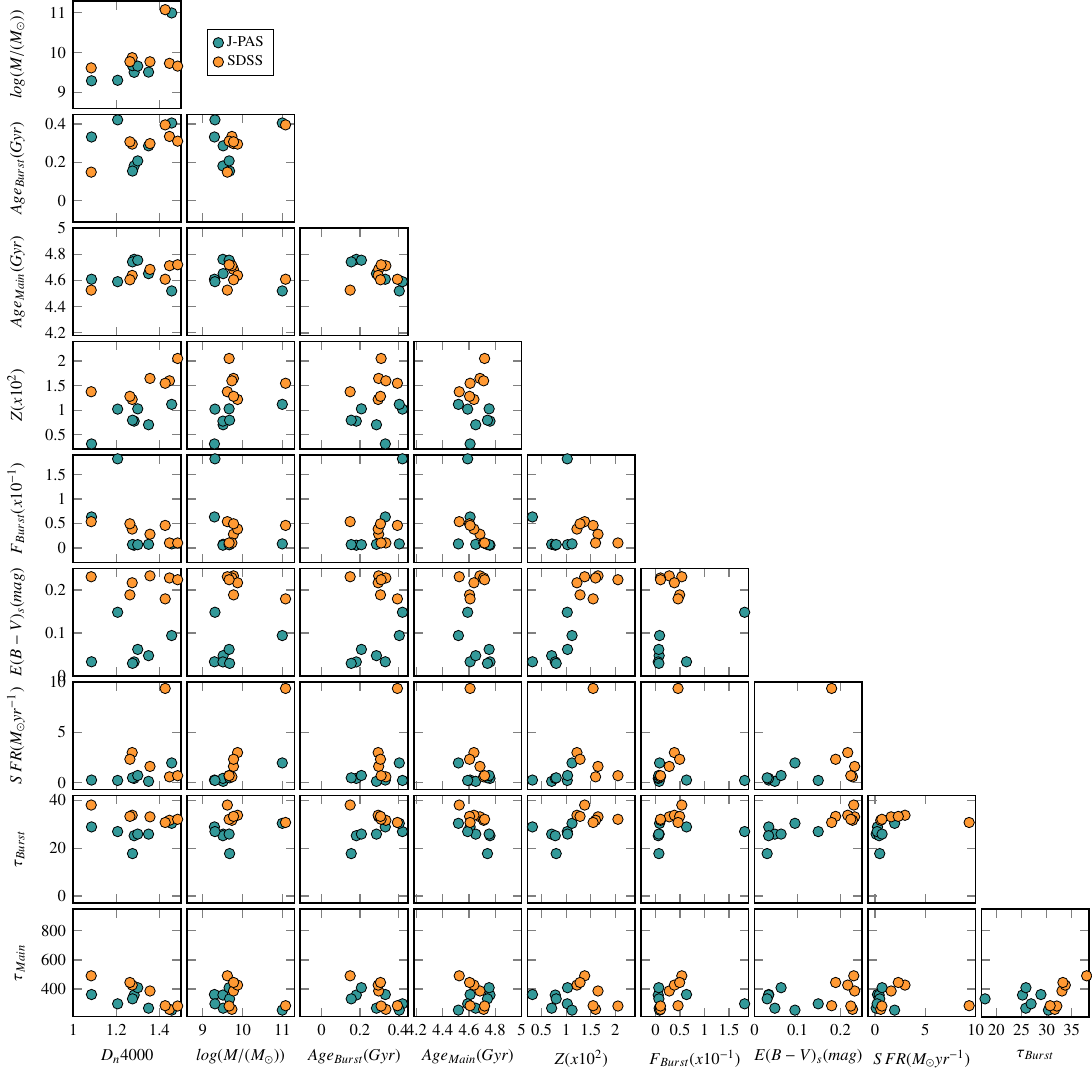}
  \fi

  \end{center}
  \caption{\label{fig:cornerplotsalpiter}
    Corner plot of CIGALE output properties for J-PAS (in cyan) and SDSS data (in orange).
    In this run, the \citet{salpeter55} IMF was used, and the main grid parameters modified in CIGALE were the same as those described in Appendix~\ref{appendix:cigaleconfigure}.
    This step was taken because it is essential to ensure that the IMF applied in the CIGALE fitting is consistent with the IMF assumed in the M/L relations adopted by B03.
  The data used to generate this figure are available on Zenodo (see Appendix~\ref{appendix:zenodocornerplotsalpiter}).}
\end{figure*}

\section{CIGALE excluding starburst}
\label{appendix:nostarburst}

\begin{figure*}[t]
  \begin{center}
  \ifdefined\makepdf%
    \tikzsetnextfilename{fig-corner-plot-no-sfr}%
    \input{tex/src/fig-corner-plot-no-sfr.tex}%
  \else
    \includegraphics[width=\linewidth]{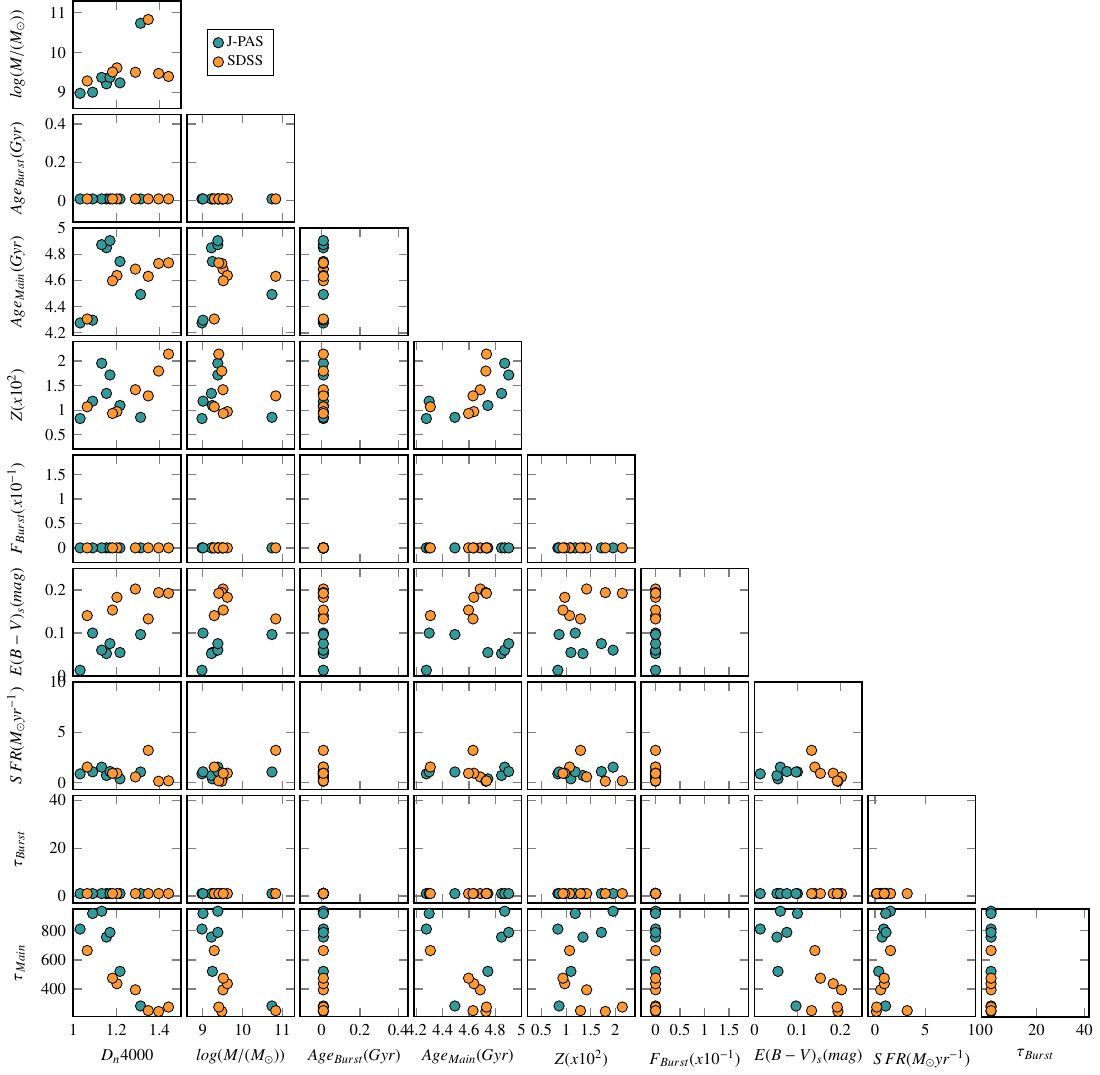}
  \fi

  \end{center}
  \caption{\label{fig:cornerplotnosfr}
    The corner plot of the CIGALE output properties for J-PAS (cyan) and SDSS (orange).
    In this analysis, both starburst and nebular components were excluded in order to isolate the underlying stellar population and reduce possible contamination in the derived parameters.
    The \citet{chabrier03} IMF was adopted, and the complete set of modified grid parameters used in this run is provided in Appendix~\ref{appendix:nostarburst}.
  The dataset used to produce this figure is publicly available on Zenodo; further details are provided in Appendix~\ref{appendix:zenodocornerplotnosfr}.}
\end{figure*}

Because we observed many emission lines in the SED fitting derived from some of the polygons with SDSS fluxes (compared with those from J-PAS in Fig.~\ref{fig:fitcigale}), and since Alba no longer shows signs of strong star formation, we decided to disable all starburst components and nebular emission in the CIGALE configuration file to see their effect on the parameters discussed here.
The resulting output properties are presented in the corner plot of Fig.~\ref{fig:cornerplotnosfr}.

By comparing Fig.~\ref{fig:cornerplotnosfr} (without starburst components) with Fig.~\ref{fig:cornerplot} (with starburst components), it is evident that the J-PAS results change significantly.
This is due to its higher resolution and larger number of filters: producing significantly different fits when the conditions change so drastically.
Whereas the SDSS results remain largely unaffected after the omission of the starburst components, likely because of its lower resolution.

The main grid parameters used without the starburst components in the CIGALE configuration file are listed below.
For additional details, the complete configuration file is available on Zenodo.

\begin{verbatim}
tau_main = 50.0,  100.0, 500.0, 1000.0
age_main = 10.,   100.,  500.,  1000., 2000.,
           3000., 4000.,
           4025., 4050., 4075., 4100., 4125.,
           4150., 4175., 4200., 4225., 4250.,
           4275., 4300., 4325., 4350., 4375.,
           4400., 4425., 4450., 4475., 4500.,
           4525., 4550., 4575., 4600., 4625.,
           4650., 4675., 4700., 4725., 4750.,
           4775., 4800., 4825., 4850., 4875.,
           4900., 4925., 4950., 4975., 5000.,
           5025., 5050., 5075., 5100., 5125.,
           5150., 5175., 5200., 5225., 5250.,
           5275., 5300., 5325., 5350., 5375.,
           5400., 5425., 5450., 5475., 5500.
tau_burst   = 1.0
age_burst   = 10.
f_burst     = 0
metallicity = 0.0001, 0.0004, 0.004, 0.008,
              0.02, 0.05
E_BV_lines  = 0.0, 0.1, 0.2, 0.3, 0.5, 0.7,
              0.8
E_BV_factor = 0.44
sfr_A       = 1.0
\end{verbatim}

\begin{figure*}[t]
  \begin{center}
  \ifdefined\makepdf%
    \tikzsetnextfilename{fig-corner-plot-new}%
    \input{tex/src/fig-corner-plot-new.tex}%
  \else
    \includegraphics[width=\linewidth]{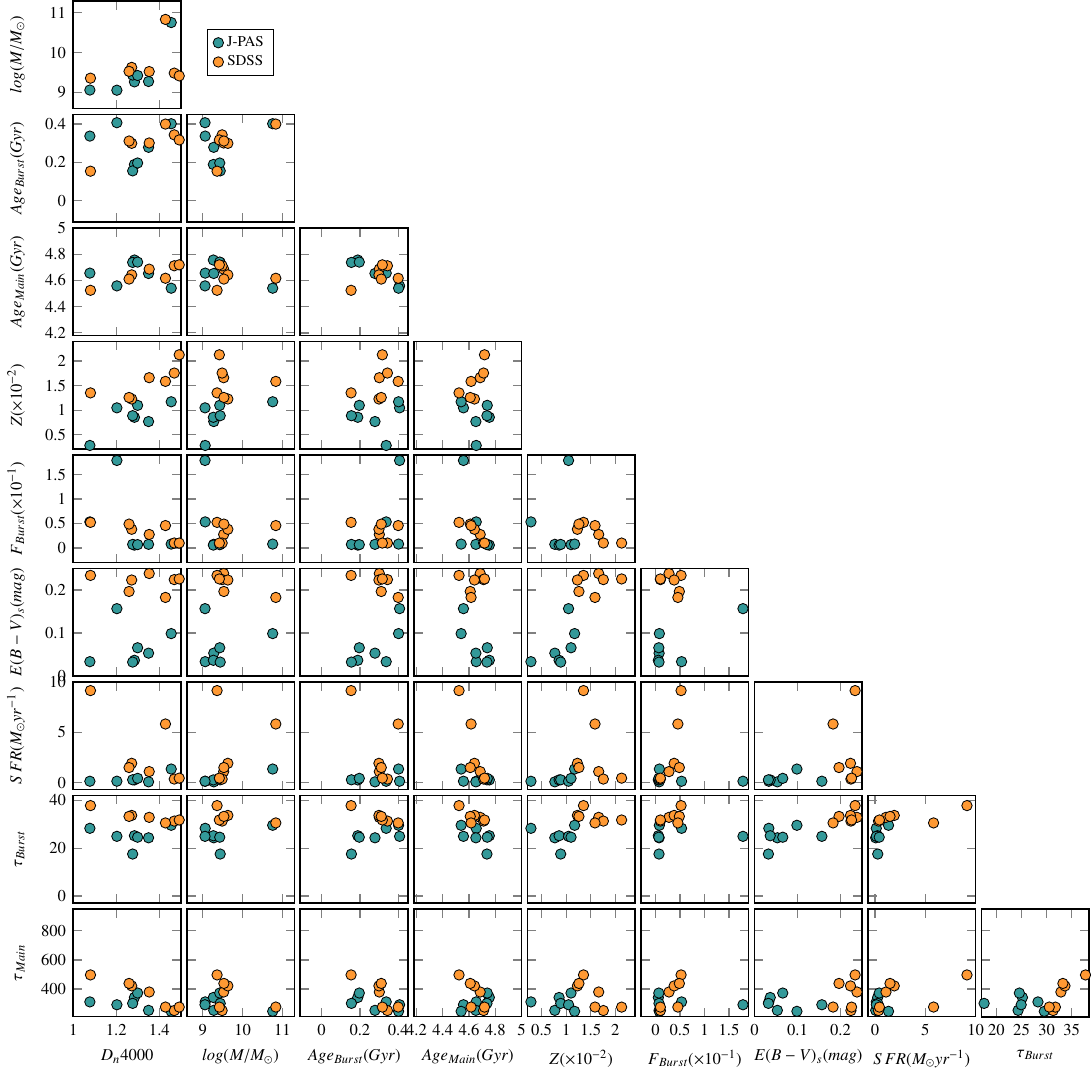}
  \fi

  \end{center}
  \caption{\label{fig:cornerplotnew}
    Same as Fig.~\ref{fig:cornerplot}, except for the ranges of the axes which are the same as Figs.~\ref{fig:cornerplotnosfr} and \ref{fig:cornerplotsalpiter}.}
\end{figure*}

Figure Fig.~\ref{fig:cornerplotnew} is identical to Fig.~\ref{fig:cornerplot}, except for the ranges of the X and Y axes.
To facilitate a clear comparison between Figs.~\ref{fig:cornerplotnosfr}, \ref{fig:cornerplotsalpiter}, and \ref{fig:cornerplotnew}, all figures have been plotted using the same axis ranges.

\section{figure and table data}
\label{appendix:zenododataforplots}
This section provides links to the archived files of this project on \href{https://doi.org/10.5281/zenodo.17674805}{zenodo.17674805}.
It contains all project source files to reproduce this analysis, as well as all the output tables behind every image/table in the PDF paper.

\subsection{Polygon coordinates (Fig.~\ref{fig:colorimg})}
\label{appendix:zenodocolorimg}
The coordinates of the polygons shown in Fig.~\ref{fig:colorimg} are provided in separate files: North-1 (\href{https://zenodo.org/records/17674805/files/polygon-1.reg}{polygon-1.reg}), North-2 (\href{https://zenodo.org/records/17674805/files/polygon-2.reg}{polygon-2.reg}), South (\href{https://zenodo.org/records/17674805/files/polygon-3.reg}{polygon-3.reg}), East (\href{https://zenodo.org/records/17674805/files/polygon-4.reg}{polygon-4.reg}), West-1 (\href{https://zenodo.org/records/17674805/files/polygon-5.reg}{polygon-5.reg}), and West-2 (\href{https://zenodo.org/records/17674805/files/polygon-6.reg}{polygon-6.reg}).

\subsection{Data used to generate Fig.~\ref{fig:manga}}
\label{appendix:zenodomanga}
The datasets used to generate the plot in Fig.~\ref{fig:manga} are available in the following files.
The gray line corresponds to \href{https://zenodo.org/records/17674805/files/manga.txt}{manga.txt}.
The orange circles in the upper panel correspond to \href{https://zenodo.org/records/17674805/files/jpas.txt}{jpas.txt}, while the teal triangles in the upper panel correspond to \href{https://zenodo.org/records/17674805/files/sdss.txt}{sdss.txt}.
The black line is provided in \href{https://zenodo.org/records/17674805/files/manga-smooth.txt}{manga-smooth.txt}.
The orange line corresponds to \href{https://zenodo.org/records/17674805/files/cigale-sed-fitting-center-jpas.txt}{cigale-sed-fitting-center-jpas.txt}, and the teal line corresponds to  \href{https://zenodo.org/records/17674805/files/cigale-sed-fitting-center-sdss.txt}{cigale-sed-fitting-center-sdss.txt}.
In the lower panel, the gray circles correspond to \href{https://zenodo.org/records/17674805/files/diff-manga-jpas.txt}{diff-manga-jpas.txt}, the teal triangles correspond to \href{https://zenodo.org/records/17674805/files/diff-obs-sed-sdss.txt}{diff-obs-sed-sdss.txt}, and the orange circles correspond to \href{https://zenodo.org/records/17674805/files/diff-obs-sed-jpas.txt}{diff-obs-sed-jpas.txt}.

\subsection{Surface brightness data for Fig.~\ref{figapen:sedpolygon}}
\label{appendix:zenodosedpolygon}
The data used to generate the Fig.~\ref{figapen:sedpolygon} can be accessed in the following files.
Surface brightness data expressed in wavelength flux density units are provided for the following regions: North-1 (\href{https://zenodo.org/records/17674805/files/polygon-1-flux-density.txt}{polygon-1-flux-density.txt}), North-2 (\href{https://zenodo.org/records/17674805/files/polygon-2-flux-density.txt}{polygon-2-flux-density.txt}), South (\href{https://zenodo.org/records/17674805/files/polygon-3-flux-density.txt}{polygon-3-flux-density.txt}), East (\href{https://zenodo.org/records/17674805/files/polygon-4-flux-density.txt}{polygon-4-flux-density.txt}), West-1 (\href{https://zenodo.org/records/17674805/files/polygon-5-flux-density.txt}{polygon-5-flux-density.txt}), West-2 (\href{https://zenodo.org/records/17674805/files/polygon-6-flux-density.txt}{polygon-6-flux-density.txt}).
Similarly, the surface brightness measurements expressed in AB magnitude units are available for the same regions: North-1 (\href{https://zenodo.org/records/17674805/files/polygon-1-sb.txt}{polygon-1-sb.txt}), North-2 (\href{https://zenodo.org/records/17674805/files/polygon-2-sb.txt}{polygon-2-sb.txt}), South (\href{https://zenodo.org/records/17674805/files/polygon-3-sb.txt}{polygon-3-sb.txt}), East (\href{https://zenodo.org/records/17674805/files/polygon-4-sb.txt}{polygon-4-sb.txt}), West-1 (\href{https://zenodo.org/records/17674805/files/polygon-5-sb.txt}{polygon-5-sb.txt}), West-2 (\href{https://zenodo.org/records/17674805/files/polygon-6-sb.txt}{polygon-6-sb.txt}).

\subsection{SED fitting data for Fig.~\ref{fig:fitcigale}}
\label{appendix:zenodofitcigal}
The data required to reproduce the plots shown in Fig.~\ref{fig:fitcigale} are available in following files.
The dark purple circles and lines corresponding to the North-1 region are provided in the files \href{https://zenodo.org/records/17674805/files/north1-jpas.txt}{north1-jpas.txt} and \href{https://zenodo.org/records/17674805/files/north1-cgal-jpas.txt}{north1-cgal-jpas.txt}, respectively.
The light purple circles and lines associated with North-2 are available in  \href{https://zenodo.org/records/17674805/files/north2-jpas.txt}{north2-jpas.txt} and \href{https://zenodo.org/records/17674805/files/north2-cgal-jpas.txt}{north2-cgal-jpas.txt}.
Similarly, the blue circles and lines representing the South region are contained in \href{https://zenodo.org/records/17674805/files/south-jpas.txt}{south-jpas.txt} and \href{https://zenodo.org/records/17674805/files/south-cgal-jpas.txt}{south-cgal-jpas.txt}.
For the East region, the green circles and lines are provided in \href{https://zenodo.org/records/17674805/files/east-jpas.txt}{east-jpas.txt} and \href{https://zenodo.org/records/17674805/files/east-cgal-jpas.txt}{east-cgal-jpas.txt}, while the red circles and lines corresponding to West-1 are available in \href{https://zenodo.org/records/17674805/files/west1-jpas.txt}{west1-jpas.txt} and \href{https://zenodo.org/records/17674805/files/west1-cgal-jpas.txt}{west1-cgal-jpas.txt}.
The peach-colored circles and lines associated with West-2 can be found in \href{https://zenodo.org/records/17674805/files/west2-jpas.txt}{west2-jpas.txt}, \href{https://zenodo.org/records/17674805/files/west2-cgal-jpas.txt}{west2-cgal-jpas.txt}.
In addition, the gray triangles and lines derived from the SDSS data are distributed as follows: North-1 (\href{https://zenodo.org/records/17674805/files/north1-sdss.txt}{north1-sdss.txt}, \href{https://zenodo.org/records/17674805/files/north1-cgal-sdss.txt}{north1-cgal-sdss.txt}),
North-2 (\href{https://zenodo.org/records/17674805/files/north2-sdss.txt}{north2-sdss.txt}, \href{https://zenodo.org/records/17674805/files/north2-cgal-sdss.txt}{north2-cgal-sdss.txt}),
South (\href{https://zenodo.org/records/17674805/files/south-sdss.txt}{south-sdss.txt}, \href{https://zenodo.org/records/17674805/files/south-cgal-sdss.txt}{south-cgal-sdss.txt}),
East (\href{https://zenodo.org/records/17674805/files/east-sdss.txt}{east-sdss.txt}, \href{https://zenodo.org/records/17674805/files/east-cgal-sdss.txt}{east-cgal-sdss.txt}),
West-1: (\href{https://zenodo.org/records/17674805/files/west1-sdss.txt}{west1-sdss.txt}, \href{https://zenodo.org/records/17674805/files/west1-cgal-sdss.txt}{west1-cgal-sdss.txt}), and
West-2 (\href{https://zenodo.org/records/17674805/files/west2-sdss.txt}{west2-sdss.txt}, \href{https://zenodo.org/records/17674805/files/west2-cgal-sdss.txt}{west2-cgal-sdss.txt}).

\subsection{Data for tables~\ref{tbl:cigalout} and \ref{tbl:criterion}}
\label{appendix:zenodotabels}
The underlying data for Tables~\ref{tbl:cigalout} and \ref{tbl:criterion} are available in the files \href{https://zenodo.org/records/17674805/files/table.tex}{table.tex} and \href{https://zenodo.org/records/17674805/files/criteria.tex}{criteria.tex}, respectively.

\subsection{Corner plot data (Fig.~\ref{fig:cornerplot})}
\label{appendix:zenodocornerplots}
The data used to generate the corner plot in Fig.~\ref{fig:cornerplot} are available in the following files.
The orange circles, corresponding to the J-PAS data, are provided in \href{https://zenodo.org/records/17674805/files/jpas-corner-plot.txt}{jpas-corner-plot.txt}, while the teal circles, representing the SDSS data, are available in \href{https://zenodo.org/records/17674805/files/sdss-corner-plot.txt}{sdss-corner-plot.txt}.

\subsection{Residual plot data (Fig.~\ref{fig:figresidual})}
\label{appendix:zenodofigresidual}
The data required to reproduce the plots shown in Fig.~\ref{fig:figresidual} are available in the following files.
For North-1, the dark violet data corresponding to J-PAS are provided in \href{https://zenodo.org/records/17674805/files/north1-residual-jpas.txt}{north1-residual-jpas.txt}, while the gray teal data corresponding to SDSS can be found in \href{https://zenodo.org/records/17674805/files/north1-residual-sdss.txt}{north1-residual-sdss.txt}.
For North-2, the light violet data corresponding to J-PAS are available in \href{https://zenodo.org/records/17674805/files/north2-residual-jpas.txt}{north2-residual-jpas.txt}, and the gray teal data corresponding to SDSS are provided in \href{https://zenodo.org/records/17674805/files/north2-residual-sdss.txt}{north2-residual-sdss.txt}.
For the South region, the blue data corresponding to J-PAS are available in \href{https://zenodo.org/records/17674805/files/south-residual-jpas.txt}{south-residual-jpas.txt}, while the gray teal data corresponding to SDSS can be found in \href{https://zenodo.org/records/17674805/files/south-residual-sdss.txt}{south-residual-sdss.txt}.
For East, the green data corresponding to J-PAS are available in \href{https://zenodo.org/records/17674805/files/east-residual-jpas.txt}{east-residual-jpas.txt}, with the gray teal data corresponding to SDSS provided in \href{https://zenodo.org/records/17674805/files/east-residual-sdss.txt}{east-residual-sdss.txt}.
For West-1, the red data corresponding to J-PAS are available in \href{https://zenodo.org/records/17674805/files/west1-residual-jpas.txt}{west1-residual-jpas.txt}, while the gray teal data corresponding to SDSS can be found in \href{https://zenodo.org/records/17674805/files/west1-residual-sdss.txt}{west1-residual-sdss.txt}.
Finally, for West-2, the peach-colored data corresponding to J-PAS are available in \href{https://zenodo.org/records/17674805/files/west2-residual-jpas.txt}{west2-residual-jpas.txt}, and the gray teal data corresponding to SDSS are provided in \href{https://zenodo.org/records/17674805/files/west2-residual-sdss.txt}{west2-residual-sdss.txt}.

\subsection{Corner plot data (Salpeter 1955 IMF, Fig.~\ref{fig:cornerplotsalpiter})}
\label{appendix:zenodocornerplotsalpiter}
The datasets used to produce the corner plot shown in Fig.~\ref{fig:cornerplotsalpiter} with the \citet{salpeter55} IMF are available in separate files.
The orange circles correspond to the J-PAS data and can be accessed in \href{https://zenodo.org/records/17674805/files/jpas-salpiter.txt}{jpas-salpiter.txt}, whereas the teal circles represent the SDSS data and are available in \href{https://zenodo.org/records/17674805/files/sdss-salpiter.txt}{sdss-salpiter.txt}.

\subsection{Corner plot data (no starburst or nebular, Fig.~\ref{fig:cornerplotnosfr})}
\label{appendix:zenodocornerplotnosfr}
The datasets used to generate the corner plot in Fig.~\ref{fig:cornerplotnosfr}, without starburst and nebular components, are provided in separate files. The orange circles, representing the J-PAS data, can be accessed in \href{https://zenodo.org/records/17674805/files/jpas-no-sfr.txt}{jpas-no-sfr.txt}, while the teal circles, corresponding to the SDSS data, are available in \href{https://zenodo.org/records/17674805/files/sdss-no-sfr.txt}{sdss-no-sfr.txt}

\section{software acknowledgment}
\label{appendix:software}
 
This research was done with the following free software programs and libraries: Bzip2 1.0.8, CFITSIO 4.6.3, cURL 8.11.1, Dash 0.5.12, Discoteq flock 0.4.0, Expat 2.6.4, File 5.46, Fontconfig 2.16.0, FreeType 2.13.3, Git 2.48.1, GNU Astronomy Utilities 0.24 \citep{gnuastro}, GNU Autoconf 2.72, GNU Automake 1.17, GNU AWK 5.3.1, GNU Bash 5.2.37, GNU Binutils 2.43.1, GNU Bison 3.8.2, GNU Compiler Collection (GCC) 15.2.0, GNU Coreutils 9.6, GNU Diffutils 3.10, GNU Findutils 4.10.0, GNU gettext 0.23.1, GNU gperf 3.1, GNU Grep 3.11, GNU Gzip 1.13, GNU Integer Set Library 0.27, GNU libiconv 1.18, GNU Libtool 2.5.4, GNU libunistring 1.3, GNU M4 1.4.19, GNU Make 4.4.1, GNU Multiple Precision Arithmetic Library 6.3.0, GNU Multiple Precision Complex library, GNU Multiple Precision Floating-Point Reliably 4.2.1, GNU Nano 8.3, GNU NCURSES 6.5, GNU Readline 8.2.13, GNU Scientific Library 2.8, GNU Sed 4.9, GNU Tar 1.35, GNU Texinfo 7.2, GNU Wget 1.25.0, GNU Which 2.23, GPL Ghostscript 10.06.0, Help2man , Less 668, Libffi 3.4.7, libICE 1.1.2, Libidn 1.42, Libjpeg 9f, Libpaper 1.1.29, Libpng 1.6.46, libpthread-stubs (Xorg) 0.5, libSM 1.2.5, Libtiff 4.7.0, libXau (Xorg) 1.0.12, libxcb (Xorg) 1.17.0, libXdmcp (Xorg) 1.1.5, libXext 1.3.6, Libxml2 2.13.5, libXt 1.3.1, Lzip 1.25, OpenSSL 3.4.0, PatchELF 0.13, Perl 5.40.1, pkg-config 0.29.2, podlators 6.0.2, Python 3.13.2, SAOImage DS9 8.7b1 \citep{ds9}, util-Linux 2.41.3, util-macros (Xorg) 1.20.2, WCSLIB 8.4, X11 library 1.8, XCB-proto (Xorg) 1.17.0, xorgproto 2024.1, xtrans (Xorg) 1.5.2, XZ Utils 5.6.3 and Zlib 1.3.1. 
The \LaTeX{} source of the paper was compiled to make the PDF using the following packages: biber 2.21, biblatex 3.21, caption 77682 (revision), courier 77161 (revision), csquotes 5.2o, datetime 2.60, fancyvrb 4.6, fmtcount 3.12, fontaxes 2.0.2, footmisc 7.0b, fp 2.1d, helvetic 77161 (revision), kastrup 15878 (revision), logreq 1.0, mweights 77682 (revision), newtx 1.756, pgf 3.1.11a, pgfplots 1.18.2, preprint 2011, setspace 6.7b, sttools 3.5, texgyre 2.501, times 77161 (revision), titlesec 2.17, trimspaces 1.1, txfonts 77682 (revision), ulem 77682 (revision), xcolor 3.02, xkeyval 2.10, xpatch 0.3 and xstring 1.86. 
We are very grateful to all their creators for freely  providing this necessary infrastructure. This research  (and many other projects) would not be possible without  them.

\end{appendix}
\end{document}